\documentclass[11pt]{article}
\usepackage[utf8]{inputenc}
\pdfoutput=1
\usepackage{setspace}
\usepackage{palatino}
\usepackage{graphicx}
\usepackage{adjustbox}
\usepackage{subfigure}
\usepackage{float}
\usepackage{titling} 
\usepackage{multirow}
\usepackage{lscape}
\usepackage{amsmath}
\usepackage{amssymb}
\usepackage{mathtools}
\usepackage{booktabs}
\usepackage{rotating}
\usepackage{amsmath}
\usepackage{pdfpages}
\usepackage{float}
\usepackage{booktabs,caption}
\usepackage[flushleft]{threeparttable}
\usepackage[utf8]{inputenc}
\usepackage{enumitem}
\usepackage{placeins}
\usepackage{breqn}
\usepackage{enumitem}
\usepackage[a4paper, total={6in, 9.5in}]{geometry}
\fontfamily{ppl}\selectfont 
\usepackage[american]{babel}
\usepackage{bm}
\usepackage{caption}
\usepackage{array}
\usepackage{booktabs}
\usepackage{dcolumn}
\newcolumntype{d}{D{.}{.}{2.5}}           
\usepackage[group-separator={,}]{siunitx}
\usepackage[normalem]{ulem}
\usepackage[textsize=small]{todonotes}

\newcommand{\q}[1]{'#1'}

\newcommand{\Cov}{\textrm{Cov}}
\newcommand{\Var}{\textrm{Var}}
\newcommand{\even}{\text{even}}
\newcommand{\odd}{\text{odd}}
\newcommand{\Poisson}{\text{Poisson}}

\usepackage[natbibapa]{apacite}
\PassOptionsToPackage{hyperindex,breaklinks}{hyperref}
\usepackage{hyperref}
\usepackage{xcolor}
\definecolor{darkblue}{rgb}{0, 0, 0.5}
\hypersetup{colorlinks=true,citecolor=darkblue, linkcolor=darkblue, urlcolor=darkblue}

\title{Bias correction in multiple-systems estimation}

\author{Daan B. Zult\\
   {\small\raggedright Statistics Netherlands}\\
   \href{mailto:db.zult@cbs.nl}{\texttt{db.zult@cbs.nl}}
\and Peter G. M. van der Heijden\\
    {\small\raggedright Utrecht University and University of Southampton}\\
    \href{mailto:P.G.M.vanderHeijden@uu.nl}{\texttt{P.G.M.vanderHeijden@uu.nl}}
\and Bart F. M. Bakker\\
    {\small\raggedright Statistics Netherlands and VU University Amsterdam}\\
    \href{mailto:bfm.bakker@cbs.nl}{\texttt{bfm.bakker@cbs.nl}}
}

\date{}
\date{\vspace{-5ex}}

\renewenvironment{abstract}
 {\par\noindent\textbf{\abstractname: }\ \ignorespaces}
 {\par\medskip}

\begin{document}

{\setstretch{.8}
\maketitle
\begin{abstract}
If part of a population is hidden but two or more sources are available that each cover parts of this population, dual- or multiple-system(s) estimation can be applied to estimate this population. For this it is common to use the log-linear model, estimated with maximum likelihood. These maximum likelihood estimates are based on a non-linear model and therefore suffer from finite-sample bias, which can be substantial in case of small samples or a small population size. This problem was recognised by Chapman, who derived an estimator with good small sample properties in case of two available sources. However, he did not derive an estimator for more than two sources. We propose an estimator that is an extension of Chapman's estimator to three or more sources and compare this estimator with other bias-reduced estimators in a simulation study. The proposed estimator performs well, and much better than the other estimators. A real data example on homelessness in the Netherlands shows that our proposed model can make a substantial difference.

\end{abstract}
}

\section{Introduction}  \label{sec:intro}

A well-known statistical problem concerns the estimation of the size of a population that is only partly observed by different sources. By linking the records in the sources the number of units observed by at least one source is found, but the number of units that are missed by all sources is unknown. The standard method to estimate this hidden number is known as dual-system estimation (DSE) for two lists and multiple-systems estimation (MSE) for more than two lists. Other names found in the literature are \textit{capture-recapture}, \textit{mark-recapture}, \textit{multiple-recapture} and \textit{multiple-record systems} estimation. A literature overview is provided by Chao (\citeyear{Chao2001}), who discusses these models in the context of human populations.

DSE leans on a set of assumptions extensively described by, for example, \cite{Wolter1986} and \cite{Zhang2019}. The \cite{IWGDMF1995} summarize them as:

\begin{enumerate}
\item There is no change in the population during the investigation (the population is closed).
\item There is no loss of tags (individuals can be linked from capture to recapture).
\item For each sample, each individual has the same chance of being included in the sample.
\item The two samples are independent. 
\end{enumerate}
Earlier Seber (\citeyear{Seber1982}) and later Chao et al. (\citeyear{Chao2001}) and van der Heijden et al. (\citeyear{vdHeijden2012}) showed that assumption 3 can be further relaxed, i.e., it is sufficient that each individual has the same chance of being included in only one of the samples, instead of both samples. In MSE, samples are allowed to be dependent, and in practical situations this makes MSE much more realistic than DSE.

Under the appropriate assumptions and conditions, a maximum likelihood (ML) estimator can be derived for the hidden and total population size. However, in finite samples these ML-estimators are mean-biased \citep[see e.g.][]{Chapman1951,Bailey1951,Rivest2001}. This mean-bias can be shown for the ML-estimators directly, but also follows more generally from the fact that these estimators make use of a hierarchical log-linear model \citep{Fienberg1972}, which provides median-unbiased, but not mean-unbiased estimates (\citealp[see e.g.][chap. 7]{Hald1952}, \citealp[or][]{Miller1984}). This finite-sample mean-bias (from now on referred to as finite-sample bias or simply bias) can be substantial in case of small samples (\citealp[p. 53-54]{Long1979}; \citealp{Rainey2021}).

The role of finite-sample bias in the discussion on the robustness and accuracy of MSE estimators is generally small. The focus is usually on other issues that lead to inaccurate estimates, such as failing model assumptions \cite[see e.g.][]{Gerritse2015,Zult2021} or model selection uncertainty \cite[see e.g.][]{Silverman2020,Binette2022}. While it is true that these issues can potentially lead to large estimation bias, it is not clear how these issues are affected by finite-sample bias, simply because it is usually ignored. This is unfortunate, because correcting for finite-sample bias comes at almost no costs to researchers \cite[]{Rainey2021}, while, as we will see, its impact can be substantial and therefore may affect conclusions.

The first to address the problem of finite-sample bias in DSE were \cite{Chapman1951} and \cite{Bailey1951}. To reduce bias, they both proposed their own bias-reduced DSE estimator. Chapman showed that his estimator is essentially unbiased \citep[][p. 145]{Chapman1951} and it became the most well-known of the two. Neither the Chapman nor Bailey estimator was extended towards MSE. The main contribution of this paper is the proposal of a Chapman MSE-estimator.

Our proposed Chapman MSE-estimator is not the first estimator that aims to reduce bias in the ML-estimator. \cite{EvansBias1994} and \cite{Rivest2001} proposed population size estimators with the same goal. Others, such as \cite{Cordeiro1991}, \cite{Firth1993} and \cite{Kosmidis2020,Kosmidis2021}, proposed bias-reduction methods for ML-estimators in log-linear models in general, which can be used in the context of MSE. In this paper we will compare the performance of these bias-reduced MSE estimators with our Chapman MSE-estimator in simulation studies.

The paper is structured as follows. Section \ref{sec:intro_DSE} discusses DSE and bias in DSE estimators. Section \ref{sec:intro_MSE} discusses MSE and a derivation of the new Chapman MSE-estimator for \textit{saturated} log-linear models, i.e., log-linear models where the number of independent parameters equals the number of counts. In Section \ref{sec:MSE_restricted} this new estimator is generalised towards a Chapman MSE-estimator that is also valid for \textit{restricted} log-linear models. In Section \ref{sec:Homeless} the new Chapman MSE-estimator is used to estimate the number of homeless people in The Netherlands. Section \ref{sec:Discussion} discusses and concludes.

\section{Dual-system estimation} \label{sec:intro_DSE}

This section discusses DSE. We first introduce notation, then Section \ref{sec:LP_LL} proceeds with the Lincoln-Peterson estimator and the log-linear model. Section \ref{sec:distributions} discusses the different distributional assumptions that underlie DSE and some of their implications. Section \ref{sec:ChapmanBailey} introduces the problem of mean-bias and gives the bias-reduced DSE estimators proposed by \cite{Chapman1951} and \cite{Bailey1951}. This section also presents an alternative interpretation of the derivation of the Chapman-estimator that has the advantage that it allows the Chapman-estimator to be easily extended towards a similar estimator for multiple sources (which we will do in Section 3). Finally, in Section \ref{sec:SimDSE}, bias-reduced DSE estimators are compared in a simple simulation study.

A description of the DSE problem starts from a population that consists of $N$ unique units that are partly observed by two sources $A$ and $B$, where the units are matched between sources. Each source is a random sample from the population, so in general not all $N$ units are observed. Each unit has an inclusion pattern that tells us in which source(s) a unit was observed. This inclusion pattern is denoted as $ab$ with $a,b=1,0$, where $a=1$ stands for \q{\textit{in the first source}} and $a=0$ for \q{\textit{not in the first source}}, and the same with $b$ for the second source. This implies that the inclusion pattern $00$ belongs to the unobserved units.

DSE uses the frequencies of occurrence of each inclusion pattern, which are simply the counts of the units with identical inclusion patterns. These counts are denoted as $n_{ab}$. A vector of the observed counts is denoted as $\mathbf{n}$, excluding the unobserved count $n_{00}$ that is unknown and to be estimated. When we sum over $a$ or $b$, we replace that subscript by \q{+}. Thus $n_{10} + n_{11} = n_{1+}$ is equal to the size of the first source, and $n_{+1}$ to the size of the second source. The total number of observed units is denoted as $n$, which allows us to write $N=n+n_{00}$. $n_{ab}$ is considered a random variable with expectation $m_{ab}$. Estimates for $N$, $m_{ab}$ and $m_{00}$ are denoted by $\hat{N}^{\text{est}}$, $\hat{m}_{ab}^{\text{est}}$ and $\hat{m}_{00}^{\text{est}}$, where the superscript \q{est} indicates the estimator that was used. These bias-reduced estimators can be obtained by using adjusted counts, that we denote as $n_{ab}^{\text{est}}$ or $\mathbf{n}^{\text{est}}$. 

\subsection {The Lincoln-Petersen estimator and the log-linear model} \label{sec:LP_LL}
The first DSE model for population size estimation was proposed by \cite{Petersen1896}, and later \cite{Lincoln1930}. It is often referred to as the Lincoln-Petersen (LP) estimator. The LP-estimator can be derived from the assumption of independence between source $A$ and $B$, which implies that the odds-ratio between source $A$ and $B$, denoted by $\theta^{AB}$, is 
\begin{align}
\theta^{AB} &= \frac{m_{11}/m_{10}}{m_{01}/m_{00}} = 1, && \label{eq:OR}
\shortintertext{which leads to}
m_{00} &= \frac{m_{10}m_{01}}{m_{11}}. \label{eq:OR_m}
\end{align}
By plugging in ML estimates for $m_{ab}$, which are simply the observed values $n_{ab}^{\text{LP}} = n_{ab}$ \citep[see e.g., ][]{BishopFienberg1975}, the LP-estimator for the missing cell is
\begin{align}
\hat{m}^{\text{LP}}_{00} &= 
\frac{n_{10}^{\text{LP}}n_{01}^{\text{LP}}}{n_{11}^{\text{LP}}} = \frac{n_{10}n_{01}}{n_{11}}, && \label{eq:LP}
\shortintertext{and the population size estimate}
\hat{N}^{\text{LP}} &= n + \hat{m}^{\text{LP}}_{00} = \frac{n_{1+}n_{+1}}{n_{11}}. \label{eq:LP_N}
\end{align}
The LP-estimator for the missing cell and for the population size are ML estimators.

\cite{Fienberg1972} shows that the LP-estimator can also be obtained from log-linear parameter estimates of the log-linear model
\begin{align}
\log \mathbb{E} \left[\mathbf{n}|\mathbf{X} \right] ={\mathbf{X}\boldsymbol{\lambda}}, && \label{eq:LL_regression}
\end{align}
with, for two sources, $\mathbf{n} = \mathbf{n}^{\text{LP}} = \left(n_{11},n_{01},n_{01}\right)^{\top}$, $\mathbf{X}=\left(\begin{array}{ccc} 1 & 1 & 1 \\ 1 & 1 & 0 \\ 1 & 0 & 1 \end{array} \right)$ and $\boldsymbol{\lambda}=\left(\lambda,\lambda_{a}^{A},\lambda_{b}^{B}\right)^{\top}$. $\lambda$ is the intercept term, and $\lambda_{a}^{A}$ and $\lambda_{b}^{B}$ are the respective inclusion parameters for source $A$ and $B$ that are identified by setting $\lambda_{0}^{A} = \lambda_{0}^{B} = 0$. It is further assumed that Eq. (\ref{eq:LL_regression}) also holds for $m_{00}$. The parameters of a log-linear model are usually estimated with ML, which for Eq. (\ref{eq:LL_regression}) gives the ML estimates $\hat{\lambda}^{\text{ML}}$, $\hat{\lambda}_{a}^{A,\text{ML}}$  and $\hat{\lambda}_{b}^{B,\text{ML}}$, which can be used to estimate $m_{00}$, i.e.:
\begin{align}
\hat{m}_{00}^{\text{ML}} = \exp \hat{\lambda}^{\text{ML}}, &&  \label{eq:m00_lambda}
\end{align}
where $\hat{m}_{00}^{\text{ML}}$ is equal to $\hat{m}_{00}^{\text{LP}}$. It is well known that ML-estimators for log-linear models are biased (\citealp[see e.g.][chap. 7]{Hald1952}, \citealp[or][]{Miller1984}), so this also holds for $\hat{m}_{00}^{\text{LP}}$.

\subsection{Distributional assumptions} \label{sec:distributions}
\cite{Chapman1951} and \cite{Bailey1951} showed that the LP-estimator can be derived as an ML-estimator, assuming that $n_{11}$ and $n_{01}$ conditional on $n_{1+}$ and $N$, follow a hypergeometric (Chapman) or binomial (Bailey) distribution. In the context of population size estimation, a hypergeometric distribution seems more fitting, because it assumes \textit{sampling without replacement}, which matches the \q{no duplicates} assumption (i) of \cite{Zhang2019}. \citeauthor{Bailey1951}(\citeyear{Bailey1951}, p. 294) was aware of this issue when he wrote \q{We shall assume that $n_{+1}$ is sufficiently small compared with $N$ for us to be able to ignore the complications of sampling without replacement}. However, later \cite{Darroch1958} argued that this choice is less obvious. He first showed that the LP-estimator can also be derived by assuming either
\begin{align*}
(n_{11},n_{10},n_{01},n_{00}) \sim & \text{Multinomial}\left(N,p_{11},p_{10},p_{01}\right)  &&
\shortintertext{or}
(n_{11},n_{10},n_{01},n_{00}) \sim& \text{Hypergeometric}\left(N,p_{11},p_{10},p_{01}\right)
\end{align*}
with $p_{ab} = m_{ab}/N$. \cite{Darroch1958} which of these distributions is the appropriate choice for any given experiment. He concludes that they lead to the same estimate $\hat{N}$ of $N$ and the same asymptotic estimate of $\Var(\hat{N})$, so the difference is notable only in higher moments. He further states that \q{In fact, if we had to generalize, we could say that the hypergeometric is likely to be appropriate when the main limiting factor on sample size is the trouble involved in marking animals and the multinomial when it is the difficulty in catching them.}. This implies that, for instance, if a population is partly observed by lists of records that contain unique record ID-codes, the multinomial seems to be the most appropriate choice. Finally, \citeauthor{Darroch1958} concludes that the multinomial distribution is capable of generalisations that the hypergeometric is unable to accommodate, an advantage that we will use in this paper.

Later, \citet[][p. 446]{BishopFienberg1975} showed that the assumption of a multinomial distribution can be replaced by
\begin{align*}
n_{ab} \sim & \Poisson \left(m_{ab}\right), &&
\shortintertext{with}
n_{00}= &N-n_{11}-n_{10}-n_{01},
\end{align*}
without loss of generality. Both the multinomial and Poisson distribution have the practical advantage that they can deal with multiple sources more easily, but the Poisson distribution has a second advantage because it allows the simplification of some derivations due to $\Cov \left(n_{ab}, n_{\neq ab} \right) = 0$ and $\Cov \left(1/n_{ab}, n_{\neq ab} \right) = 0$.

\subsection{Bias reduction in dual-system estimation} \label{sec:ChapmanBailey}
\cite{Chapman1951} and \cite{Bailey1951} were the first to be aware of the bias in the LP-estimator. This bias can be easily seen when we assume $n_{ab} \sim \Poisson \left(m_{ab}\right)$ and write the expectation of the LP-estimator as
\begin{align} \mathbf{E}\left[\hat{m}_{00}^{\text{LP}}\right] = \mathbf{E}\left[ \frac{n_{10}n_{01}}{n_{11}}\right] = m_{10}m_{01}\mathbf{E}\left[ \frac{1}{n_{11}}\right], && \label{eq:bias_source}
\end{align}
which is not equal to $\frac{m_{10}m_{01}}{m_{11}}$ because $\mathbb{E} \left[\frac{1}{n_{11}} \right] \neq \frac{1}{m_{11}}$. This shows that under a Poisson distribution, $\frac{1}{n_{11}}$ is the only source of bias in the LP-estimator.

\citeauthor{Chapman1951} and \citeauthor{Bailey1951} started with the hypergeometric and binomial distribution respectively and used different approximation approaches of the expectation of the ML-estimator to derive their bias-reduced estimators. \citeauthor{Bailey1951} used a second-order Taylor series approximation and concludes that
\begin{align}
\hat{m}^{\text{Bailey}}_{00} &= \frac{n_{10}^{\text{Bailey}} n_{01}^{\text{Bailey}}}{n_{11}^{\text{Bailey}}} = \frac{n_{10}(n_{01}-1)}{(n_{11}+1)}  && \label{eq:Bailey}
\shortintertext{and}
\hat{N}^{\text{Bailey}} &= \frac{n_{1+}(n_{+1}+1)}{(n_{11}+1)}.\label{eq:Bailey_N}
\end{align}
are biased reduced estimators for $m_{00}$ and $N$ respectively \citep[][p. 295]{Bailey1951}.

\citeauthor{Chapman1951} uses a different approach that is recommended by \cite{Stephan1945}. Instead of a Taylor approximation, \citeauthor{Stephan1945} recommends writing $\mathbf{E}\left[ \frac{1}{x}\right]$, with $x$ a binomial random variable, as a series of inverse factorials, as one needs quite a few terms before a Taylor series becomes reasonably accurate \cite[][p. 52]{Stephan1945}. This increased rate of convergence of \citeauthor{Stephan1945}'s inverse factorial approximation in case of $\mathbf{E}\left[ \frac{1}{x}\right]$ and $n_{11} \sim \Poisson (m_{11})$, is illustrated with a straightforward simulation study presented in Appendix \ref{App:Taylor_vs_IF}. \citeauthor{Chapman1951} uses \citeauthor{Stephan1945}'s inverse factorial approximation to derive a bias-reduced expression for $\frac{n_{10}n_{01}}{n_{11}}$ and concludes that a bias-reduced estimator for $m_{00}$ is
\begin{align}
\hat{m}^{\text{Chap}}_{00} &= \frac{n_{10}^{\text{Chap}} n_{01}^{\text{Chap}}}{n_{11}^{\text{Chap}}} = \frac{n_{10}n_{01}}{(n_{11}+1)},  && \label{eq:Chap}
\shortintertext{and for $N$}
\hat{N}^{\text{Chap}} &= \frac{(n_{1+}+1)(n_{+1}+1)}{(n_{11}+1)}-1. \label{eq:Chap_N}
\end{align}
A Bailey and Chapman estimate can also be obtained from the log-linear model in Eq. (\ref{eq:LL_regression}), if instead of $\mathbf{n}^{\text{LP}} = \left(n_{11},n_{10},n_{01}\right)^{\top}$, respectively, $\mathbf{n}^{\text{Bailey}} = \left(n_{11}+1,n_{10},n_{01} -1 \right)^{\top}$ and
$\mathbf{n}^{\text{Chap}} = \left(n_{11}+1,n_{10},n_{01}\right)^{\top}$ are used.

The Chapman- and Bailey-estimators differ only slightly, but the Chapman-estimator became the standard bias-reduced estimator in DSE literature. A reason could be that \citet[][p. 146]{Chapman1951} further shows that if $\frac{n_{1+}n_{+1}}{N}>\log \left( {\frac{N}{\epsilon}} \right)$ holds,
\begin{align*}
\left| \mathbb{E} \left[\hat{N}^{\text{Chap}} \right] - N \right| < \frac{\epsilon}{100} N, &&
\end{align*}
with $\epsilon$ some arbitrary small positive number \citep[][p. 502]{Cramer1922} also holds. This means that if the two sources are large enough compared to $N$, the bias in $\hat{N}^{\text{Chap}}$ is less than $\epsilon$ percent of $N$ and so Chapman refers to his estimator as \q{essentially unbiased}. Therefore we refer to the Chapman-estimator not only as a bias-reduced, but also as a bias-\textit{corrected} estimator. Chapman, (\citeyear{Chapman1951}, p. 146) finally notes that the Chapman-estimator requires 
\begin{align} \label{eq:regularity_DSE}
\frac{n_{1+}n_{+1}}{N}>\log N, &&
\end{align}
to hold. This inequality is derived from setting $\left|\mathbf{E}\left[\hat{N}^{\text{Chap}}\right]-N\right| \le 1$ and can be considered a regularity condition for the Chapman-estimator. If this regularity condition is not met, $\hat{N}^{\text{Chap}}$ may suffer from considerable (negative) bias, as we will illustrate later in the simulation study in Table \ref{tab:DSE}.

Chapman derived his estimator for the hypergeometric distribution, but it can also be developed with  a second-order Taylor approximation for the multinomial and Poisson distribution, which is derived in Appendix \ref{App:BiasinDSE}. This derivation suggests that the Chapman-estimator is also valid under a multinomial or Poisson distribution. This is useful when we extend the Chapman-estimator to multiple sources in Section \ref{sec:ChapmanMSE_saturated}. Combining the Chapman-estimator with the results in Appendix \ref{App:Taylor_vs_IF} and \ref{App:BiasinDSE} imply that if $n_{ab} \sim \Poisson(m_{ab})$ we can write
\begin{align} \label{eq:correction}
\frac{1}{m_{ab}} \approx \mathbf{E}\left[\frac{1}{n_{ab}+1}\right]. &&
\end{align}
This equation will allow us to easily extend the Chapman MSE-estimator towards multiple sources in Section \ref{sec:ChapmanMSE_saturated}.

\citeauthor{Bailey1951} did not extend his estimator to more than two sources. \citeauthor{Chapman1952} (\citeyear{Chapman1952}) did, but he only considered the case where a unit was tagged in an earlier source or not, and did not consider dependence between pairs of sources. Dependence between sources is further discussed in Section \ref{sec:MSEbasics}.
Others, like \cite{Cordeiro1991}, \cite{Firth1993}, \cite{EvansBias1994}, \cite{Rivest2001}, \cite{Kosmidis2020}, have proposed bias-reduced estimators for log-linear models in general and therefore  do take dependence between sources into account. These models are discussed in more detail in Section \ref{sec:BiasReducedMSE}. However, we will include these bias-reduced estimators in the simple DSE simulation study presented in Section \ref{sec:SimDSE}. 

\subsection{Dual-system estimation simulation study} \label{sec:SimDSE}
In this section we compare the LP-, Bailey-, Cordeiro, Firth-, Kosmidis, Evans and Bonette (EB)-, Rivest and Lévesque (RL)- and Chapman-estimator in a DSE setting. The LP-, Bailey- and Chapman-estimator can only be used in DSE and were discussed in the previous sections. The Cordeiro, Firth-, Kosmidis, EB- and RL-estimator can be applied in both DSE and MSE and will be discussed in Section \ref{sec:BiasReducedMSE}. We use a Monte Carlo simulation study to compare the different estimators. The method we use to generate contingency tables is discussed in \cite{Hammond2023}. It allows us to start with a log-linear model having prespecified inclusion probabilities $p_{A}$ and $p_{B}$ and odds ratio(s) and generate contingency tables from this model. The resulting $n_{ab}$ are generated from a multinomial distribution. This is particularly useful in the next section in which we consider more than two sources, and pairs of sources that are dependent.

A minor but important simulation issue is the regularity condition in Eq. (\ref{eq:regularity_DSE}), or the issue of what \citet[][p. 125]{Otis1978} refer to as \q{failures}. This implies that the relation between $n_{1+}$, $n_{+1}$ and $N$ must be set such that they comply with Eq. (\ref{eq:regularity_DSE}). A simple example of a failure is when, in DSE, $n_{11}$ equals zero, which leads to $\hat{N}^{\text{LP}} = \infty$. \cite{Otis1978} recommend replacing such a replication with a new replication, an advice that was followed in \cite{EvansBias1994}. However, replacing failure replications, that correspond to large population size estimates, with new population size estimates, introduces selection bias in the sense that, when $\hat{N}^{\text{est}}$ is an unbiased estimator for $N$, the mean of these estimates $\bar{\hat{N}}^{\text{est}}$ departs from $N$. Therefore, to obtain accurate mean estimates that allow a fair comparison of bias between the different estimators, we choose the combined $N$, $p_{A}$ and $p_{B}$ such that for scenario $1-6$ the probability of failures becomes close to zero. Nonetheless a failure occurred once for scenario $1$. These settings also imply that the regularity condition in Eq. (\ref{eq:regularity_DSE}) holds by a substantial margin.

The scenario parameters are shown in the columns $N, p_A$ and $p_B$ of Table \ref{tab:DSE} below. Scenarios 1 - 6 comply to Chapman's regularity condition in Eq. (\ref{eq:regularity_DSE}). To see how estimators are affected when this regularity condition is violated, we have added a $7^{th}$ scenario under which the regularity condition does not hold, i.e. $n_{1+} = n_{+1} = 15$, so $n_{1+}n_{+1}/N = 225/100 < \log 100$. The different estimators that are compared are shown in the columns that follow. In the context of DSE some estimators are equivalent and their results are displayed in a single column. This holds for $\hat{N}^{\text{EB}}$, $\hat{N}^{\text{Cordeiro}}$, $\hat{N}^{\text{Firth}}$ and $\hat{N}^{\text{Kosmidis}}$ (denoted as $\hat{N}^{\text{EB/CFK}}$), and for $\hat{N}^{\text{Chap}}$ and $\hat{N}^{\text{RL}}$ (denoted as $\hat{N}^{\text{Chap/RL}}$).

\begin{table}[htbp]
  \caption{Simulation study with $20,000$ replications for seven DSE scenarios.}\label{tab:DSE}
\begin{tabular}{l|rr@{}lr@{}lr@{}l|r@{}lr@{}lr@{}lr@{}lr@{}lr@{}l}
\toprule
$S$ & $N$ & \multicolumn{2}{c}{$p_{A}$} & \multicolumn{2}{c}{$p_{B}$} & \multicolumn{2}{c|}{$\bar{n}$} & \multicolumn{2}{c}{$\bar{\hat{N}}^{\text{LP}}$} & \multicolumn{2}{c}{$\bar{\hat{N}}^{\text{Bailey}}$} & \multicolumn{2}{c}{$\bar{\hat{N}}^{\text{EB/CFK}}$} & \multicolumn{2}{c}{$\bar{\hat{N}}^{\text{Chap/RL}}$} \\
\midrule
1 & 100 & 0&.5 & 0&.2 &60&.0 & 105&.3$^{\ast\ast\ast\dagger}$ &96&.1$^{\ast\ast\ast}$ & 105&.2$^{\ast\ast\ast}$ & 100&.1 \\
2 & 100 & 0&.35 & 0&.3 &54&.5 & 106&.0$^{\ast\ast\ast}$ &98&.0$^{\ast\ast\ast}$ & 105&.3$^{\ast\ast\ast}$ & 100&.4$^{\ast}$ \\
3 & 500 & 0&.4 & 0&.15 & 244&.9 & 508&.3$^{\ast\ast\ast}$ & 493&.6$^{\ast\ast\ast}$ & 507&.4$^{\ast\ast\ast}$ & 499&.2 \\
4 & 500 & 0&.25 & 0&.2 & 200&.1 & 512&.4$^{\ast\ast\ast}$ & 495&.4$^{\ast\ast\ast}$ & 509&.3$^{\ast\ast\ast}$ & 499&.4 \\
5 & 10,000 & 0&.3 & 0&.1 &3,699&.2 & 10,018&.0$^{\ast\ast\ast}$ &9,987&.9$^{\ast\ast\ast}$ & 10,013&.1$^{\ast\ast\ast}$ &9,996&.9 \\
6 & 10,000 & 0&.25 & 0&.15 &3,624&.9 & 10,016&.7$^{\ast\ast\ast}$ &9,993&.9$^{\ast}$ & 10,012&.5$^{\ast\ast\ast}$ &9,999&.6 \\
7 & 100 & 0&.15 & 0&.15 &27&.8 & 146&.2$^{\ast\ast\ast\dagger}$ &87&.2$^{\ast\ast\ast}$ & 128&.0$^{\ast\ast\ast}$ &92&.3$^{\ast\ast\ast}$ \\
\bottomrule
\end{tabular}
 \begin{tablenotes}
      \small
      \item       
      $\bar{n}$ gives the mean number of observed units $n$ over all replications. The superscripts $^{\ast}$, $^{\ast\ast}$ and $^{\ast\ast\ast}$ indicate that we can reject $\hat{N}^{\text{est}}=N$ with a two-sided t-test with p-values = $0.05, 0.01$ and $0.001$ respectively. A $\dagger$ as superscript indicates that extremely high estimates due to failures were replaced with the highest Chapman estimate in the simulation sample.
 \end{tablenotes}
\end{table}
The $^{\ast}$s in the column of $\bar{\hat{N}}^{\text{Chap/RL}}$ indicate that for p-value $=0.05$, in five out of the six regular scenarios, the hypothesis $N = \hat{N}^{\text{Chap/RL}}$ cannot be rejected. For p-value $=0.01$ this holds for all six regular scenarios. The same does not hold for the other estimators, for which the mean over all replications, in most cases, significantly differs from $N$ for p-value $=0.001$, and for all cases for p-value $=0.05$. For all scenarios the bias in $\hat{N}^{\text{Chap/RL}}$ is smaller than the bias in the other estimators. This shows that in DSE, the Chapman- and RL-estimator are superior to the other estimators. If Chapman's regularity condition in Eq. (\ref{eq:regularity_DSE}) is not met, as in scenario 7, all estimators are considerably biased.

The standard deviation(SD) and root mean squared errors (RMSEs) that correspond to each estimator and scenario in Table \ref{tab:DSE} can be found in Table \ref{tab:DSE_RMSE} in Appendix \ref{App:sim_DSE}. This table shows that the SDs and RMSEs of the Bailey- and Chapman/RL-estimator are smaller than the RMSEs of the EB/CFK-estimator, which in turn are smaller than the RMSEs of the ML-estimator. 

\section{Multiple-systems estimation} \label{sec:intro_MSE}
This section discusses multiple-systems estimation (MSE). First it introduces some notation additional to the DSE notation introduced in Section \ref{sec:intro_DSE}. Next, Section \ref{sec:MSEbasics} proceeds with some MSE preliminaries and bias-reduced MSE estimators. In Section \ref{sec:ChapmanMSE_saturated} we derive a new bias-corrected estimator that can be considered an extension of the Chapman-estimator towards MSE under saturated models. In Section \ref{sec:MSE_restricted} the Chapman MSE-estimator is further generalised towards all log-linear models, both saturated and restricted.

MSE considers the case where a population that consists of $N$ unique units is partly observed by a set of $k$ sources, indicated by $A,B,C,...$. For ease of notation we will, where possible, discuss MSE from the perspective of three sources, because it can often be generalised to $k$ sources in a straightforward way. For three sources, the inclusion pattern is denoted as $abc$ with $a,b,c=1,0$, with the same meaning as $ab$ in DSE notation. For $k$ sources the inclusion pattern is $ab \ldots k$. We introduce notation that allows us to distinguish between the sets of unit counts that are observed an \textit{even} and \textit{odd} number of times, that we denote by $n_{\even}$ (or $m_{\even}$, $\hat{m}_{\even}$) and $n_{\odd}$ (or $m_{\odd}$, $\hat{m}_{\odd}$). For three sources this gives $n_{\odd} = \left(n_{111},n_{100},n_{010},n_{001}\right)$ and $n_{\even} = \left(n_{110},n_{101},n_{011}\right)$.

In contrast to DSE, in MSE the log-linear model can take different forms. Therefore, the superscript in $\hat{N}^{\text{est}}$, $n_{ab}^{\text{est}}$ and $\hat{m}_{ab}^{\text{est}}$ is extended to $\hat{N}^{\text{est,LLM}}$, $n_{abc}^{\text{est,LLM}}$ and $\hat{m}_{abc}^{\text{est,LLM}}$, where \q{est,LLM} specifies not only the chosen estimator but also the chosen log-linear model.

\subsection {Preliminaries} \label{sec:MSEbasics}
The first to consider more than two sources was \cite{Schnabel1938}. After this the use of multiple sources became more common and estimators were introduced that made use of different distributional assumptions. For instance, \cite{Chapman1954}, \cite{Darroch1958} and \cite{Cormack1991} assumed every element in $n_{abc}$ to be an independent realisation from a Poisson distribution. This is a reasonable assumption when $n_{abc}$ are relatively small compared to $N$, but when this is not true, one should take into account that each $m_{abc}$ has an upper-bound of $N$. Adding this restriction to the Poisson distribution assumption is equivalent to assuming that the joint set of $n_{abc}$ has a multinomial distribution with expectations $m_{abc}$ for which $m_{000} + \sum_{abc} m_{abc} = N$ \citep[see e.g.][]{Sanathanan1972,BishopFienberg1975,Wolter1986,Darroch1993}. 

In case of three sources the independence assumption that holds in DSE is relaxed and it is sufficient to assume that two conditional odds-ratios given the levels of the third source are equal. For example, for the two odds-ratios of source $A$ and $B$ given source $C$
\begin{align}
    &\frac{m_{110}/m_{100}}{m_{010}/m_{000}}= \frac{m_{111}/m_{101}}{m_{011}/m_{001}},  &&  \label{eq:OR3}
\shortintertext{which gives}
&m_{000}=\frac{m_{111}m_{100}m_{010}m_{001}}{m_{110}m_{101}m_{011}}.  && \label{eq:OR3_m000_exp}
\end{align}
A general expression is provided by \cite{Fienberg1972}, who states that for $k$ sources, $m_{00 \ldots 0}$ can be written as
\begin{align}
m_{00 \ldots 0} = \frac{\prod{m_{\odd}}}{\prod{m_{\even}}}.  && \label{eq:OR_m00...0_exp}
\end{align}
For three sources, the saturated (SAT) log-linear model for the seven observed counts becomes
\begin{align} \label{eq:LL3_SAT}
\text{SAT: } \log m_{abc} = \lambda + \lambda_{a}^{A} + \lambda_{b}^{B} + \lambda_{c}^{C} + \lambda_{ab}^{AB} +\lambda_{ac}^{AC} + \lambda_{bc}^{BC}, &&
\end{align}
where the parameters are identified by setting them to zero if one or more of the subscripts are $0$. In comparison to DSE, in the saturated log-linear model the independence assumption is replaced by the assumption of no three-factor interaction, i.e  $\lambda_{abc}^{ABC} = 0$. The interaction parameters $\lambda_{ab}^{AB}$, $\lambda_{ac}^{AC}$ and $\lambda_{bc}^{BC}$ allow for interactions between pairs of sources, and thus the model is less restrictive than the DSE model and hence more realistic in applications.

For three sources, the saturated model is not the only log-linear model that can be used. If the parameters of one or more pairs of sources are set to zero (e.g. $\lambda_{ab}^{AB} = 0$), we have a \textit{restricted} log-linear model. An advantage of further restricted models is that the resulting estimates have smaller variance than estimates from less restricted models \citep[][p. 242]{BishopFienberg1975}. A disadvantage is that they give biased estimates if the assumed restriction does not hold. We discuss restricted models in more detail because, as we will see in Section \ref{sec:MSE_restricted}, the precise model specification affects the bias-corrected estimator. \cite{Fienberg1972} and \cite{BishopFienberg1975} discuss the three possible alternative log-linear model formulations for three sources where all direct inclusion parameters $\lambda_{a}$, $\lambda_{b}$ and $\lambda_{c}$ are included. Starting from the saturated log-linear model SAT in Eq. (\ref{eq:LL3_SAT}), they discuss the two-pair dependence (2PD), the one-pair dependence (1PD), and independence (IND) model. Examples of 2PD and 1PD are
\begin{align}
& \text{2PD}: \log m_{abc} = \lambda + \lambda_{a}^{A} + \lambda_{b}^{B} + \lambda_{c}^{C} + \lambda_{ab}^{AB} + \lambda_{bc}^{BC},  && \label{eq:LL3_2PD} \\
& \text{1PD}: \log m_{abc} = \lambda + \lambda_{a}^{A} + \lambda_{b}^{B} + \lambda_{c}^{C} + \lambda_{ab}^{AB}\label{eq:LL3_1PD}
\shortintertext{and} 
& \text{IND}: \log m_{abc} = \lambda + \lambda_{a}^{A} + \lambda_{b}^{B} + \lambda_{c}^{C}. \label{eq:LL3_IND}
\end{align}

It suits our purpose to write these models as in Eq. (\ref{eq:LL_regression}). They all use $\mathbf{n}=\mathbf{n}^{\text{ML}}=\left(n_{111},n_{110},n_{101},n_{011},n_{100},n_{010},n_{001}\right)$, but differ with respect to $\boldsymbol{\lambda}=\boldsymbol{\lambda}^{\text{LLM}}$ and $\mathbf{X}=\mathbf{X}^{\text{LLM}}$. $\boldsymbol{\lambda}^{\text{LLM}}$ simply consists of the $\lambda$'s in the corresponding LLM and $\mathbf{X}^{\text{LLM}}$ becomes $X_{abc}^{\text{SAT}}$, $X_{abc}^{\text{2PD}}$, $X_{abc}^{\text{1PD}}$ or $X_{abc}^{\text{IND}}$ written as
\\
\\
\noindent ${\begin{pmatrix} 1 & 1 & 1 & 1 & 1 & 1 & 1 \\ 1 & 1 & 1 & 0 & 1 & 0 & 0 \\ 1 & 1 & 0 & 1 & 0 & 1 & 0 \\ 1 & 0 & 1 & 1 & 0 & 0 & 1 \\ 1 & 1 & 0 & 0 & 0 & 0 & 0 \\ 1 & 0 & 1 & 0 & 0 & 0 & 0 \\ 1 & 0 & 0 & 1 & 0 & 0 & 0 \end{pmatrix}}$, ${\begin{pmatrix} 1 & 1 & 1 & 1 & 1 & 1 \\ 1 & 1 & 1 & 0 & 1 & 0 \\ 1 & 1 & 0 & 1 & 0 & 0 \\ 1 & 0 & 1 & 1 & 0 & 1\\ 1 & 1 & 0 & 0 & 0 & 0 \\ 1 & 0 & 1 & 0 & 0 & 0 \\ 1 & 0 & 0 & 1 & 0 & 0 \end{pmatrix}}$, ${\begin{pmatrix} 1 & 1 & 1 & 1 & 1 \\ 1 & 1 & 1 & 0 & 1 \\ 1 & 1 & 0 & 1 & 0 \\ 1 & 0 & 1 & 1 & 0 \\ 1 & 1 & 0 & 0 & 0 \\ 1 & 0 & 1 & 0 & 0 \\ 1 & 0 & 0 & 1 & 0  \end{pmatrix}}$ or ${\begin{pmatrix} 1 & 1 & 1 & 1 \\ 1 & 1 & 1 & 0 \\ 1 & 1 & 0 & 1\\ 1 & 0 & 1 & 1 \\ 1 & 1 & 0 & 0 \\ 1 & 0 & 1 & 0 \\ 1 & 0 & 0 & 1 \end{pmatrix}}$, \\ 
\\
respectively. Estimating one of these models with ML gives the fitted-values $\hat{m}_{abc}^{\text{ML,LLM}}$.

A general expression for $m_{000}^{\text{ML,LLM}}$ is obtained by replacing the $m_{abc}$ in Eq. (\ref{eq:OR3_m000_exp}) with the ML estimates $\hat{m}_{abc}^{\text{ML,LLM}}$. For LLM = SAT this gives $m_{abc}^{\text{ML,LLM}}=m_{abc}^{\text{ML,SAT}}=n_{abc}$ and thus
\begin{align}  \label{eq:m000_OR_SAT}
\hat{m}_{000}^{\text{ML,SAT}} = \frac{n_{111}n_{100}n_{010}n_{001}}{n_{110}n_{101}n_{011}}. &&
\end{align}
For model 2PD and 1PD \citet[p. 596]{Fienberg1972} shows this expression can be further simplified, i.e.,
\begin{align}
& \hat{m}_{000}^{\text{ML,2PD}} = \frac{n_{100}n_{001}}{n_{101}}  && \label{eq:m000_2pairDependence} \shortintertext{and}
& \hat{m}_{000}^{\text{ML,1PD}} = \frac{n_{001}n_{++0}}{n_{111}+n_{101}+n_{011}}. \label{eq:m000_1pairDependence}
\end{align}
For model IND, such a closed form solution does not exist. However, for IND an estimate can be obtained by replacing the $m_{abc}$ in Eq. (\ref{eq:OR3_m000_exp}) with the fitted-values $\hat{m}_{abc}^{\text{ML,IND}}$. \cite[p. 597]{Fienberg1972} shows that this is in fact an approach that can be generally used, where Eq. (\ref{eq:OR_m00...0_exp}) gives an estimate of the missing cell for any log-linear model with any number of sources. Note that the LP-estimator in Eq. (\ref{eq:LP}) can be considered a special of (\ref{eq:OR_m00...0_exp}), with $k=2$ and LLM=SAT.

\subsubsection{Bias reduction in multiple-systems estimation} \label{sec:BiasReducedMSE}
MSE is based on a log-linear model estimated with ML, which is well known to give biased estimates (\citealp[see e.g.][chap. 7]{Hald1952}, \citealp[or][]{Miller1984}). In this context, an approach to obtain bias-reduced ML estimates is also known as generalized linear models (GLMs) using adjusted score functions \cite[see e.g.][]{Firth1993,Kosmidis2020}. A DSE example of this approach can be found in Section \ref{sec:ChapmanBailey}, where replacing $\mathbf{n}$ with the adjusted $\mathbf{n}^{\text{Chap}}$ or $\mathbf{n}^{\text{Bailey}}$ led to the bias-reduced DSE estimators by Chapman and Bailey. \cite{Cordeiro1991,Firth1993,Kosmidis2011} and others use this approach to obtain bias-reduced ML-estimators for generalised linear models, such as the log-linear model, that can also be applied to MSE. In literature we also find bias-reduced MSE estimators that are developed specifically in the context of MSE, such as by \cite{EvansBias1994} and \cite{Rivest2001}. However, despite that they are developed in the more specific MSE context, these bias-reduced MSE estimators also fit what \cite{Firth1993} refers to as the more general (GLMs with) modified-score functions approach.

The simplest bias-reduced MSE estimator is proposed by \cite{EvansBias1994}, which we denote as $\hat{N}^{\text{EB,LLM}}$. They propose to use the adjusted $\mathbf{n}^{\text{EB}}=\mathbf{n}+0.5^{(k-1)}$ in (\ref{eq:LL_regression}) instead of $\mathbf{n}$. This is the result of a compromise between \cite{Berkson1955} and \cite{Plackett1981}. In a log-linear regression model, \citeauthor{Berkson1955} proposes to replace values in $n_{abc}$ that are equal to zero with $0.5^{\left(k-1\right)}$, and \cite{Plackett1981}, who states that if $n_{abc} \sim \Poisson(m_{abc})$, then $\log\left(n_{abc}+0.5\right)$ instead of $\log\left(n_{abc}\right)$ is a more accurate estimate for $\log\left(m_{abc}\right)$.

Another bias-reduced estimator that was developed specifically for MSE was proposed by \citet[RL,][]{Rivest2001}. They propose a bias reduction method that can be used to reduce bias in a set of MSE estimators proposed by \cite{Otis1978} in the context of wildlife populations. Unfortunately, with the exception of the independence model, which corresponds to the $M_{t}$ model, the other models by \citeauthor{Otis1978} do not correspond exactly to Eq. (\ref{eq:LL3_SAT}) - (\ref{eq:LL3_1PD}). For the SAT, 2PD and 1PD model, we consider the adjusted counts that belong to model $M_{th}$ as the most appropriate choice, because it is most similar. See \citet{EvansGeneral1994} for further discussion on this topic. The bias reduction by \citeauthor{Rivest2001} is derived from a standard result by \cite{McCullagh1989}, about the bias in estimators in generalized linear models. \citeauthor{McCullagh1989} derive an asymptotic bias expression for estimates based on models with canonical link functions, such as the log-linear model. For two sources, the resulting RL-estimator is equal to the Chapman-estimator, as was seen in Table \ref{tab:DSE}. We denote the RL-estimator as $\hat{N}^{\text{RL,LLM}}$ and their adjusted count as $\mathbf{n}^{\text{RL,LLM}}$, with $\mathbf{n}^{\text{RL,IND}} = \mathbf{n}^{\text{RL,}M_{t}}$ and $\mathbf{n}^{\text{RL,SAT/2PD/1PD}} = \mathbf{n}^{\text{RL,}M_{th}}$. For three sources they become \citep[][p. 562]{Rivest2001}:
\begin{align}
\mathbf{n}^{\text{RL,}M_{t}} &= (n_{111}, n_{110}+\frac{1}{3}, n_{101}+\frac{1}{3}, n_{011}+\frac{1}{3},n_{100}+\frac{1}{6},n_{010}+\frac{1}{6},n_{001}+\frac{1}{6})^{\top} && \label{eq:Rivest_IND}
\shortintertext{and}
\mathbf{n}^{\text{RL,}M_{th}} &= (n_{111},n_{110}+\frac{1}{3},n_{101}+\frac{1}{3}, n_{011} + \frac{1}{3}, n_{100},n_{010},n_{001})^{\top}. \label{eq:Rivest_other}
\end{align}
\cite{Rivest2001} show in a simulation study that their estimator outperforms the EB-estimator in terms of bias reduction.

Bias reduction in MSE by means of the modified-score functions approach \citep{Firth1993} relies on the same standard result about the bias of estimators in GLMs by \cite{McCullagh1989} as was used by \cite{Rivest2001}. It was used by \cite{Cordeiro1991,Firth1993,Kosmidis2007,Kosmidis2011} and others to reduce bias in parameter estimates in log-linear models. \citeauthor{Cordeiro1991}, \citeauthor{Firth1993} and \citeauthor{Kosmidis2011} give three different bias-reduced parameter estimates $\hat{\lambda}^{\text{est}}$ for the $\lambda$ in Eq. (\ref{eq:LL_regression}), which correspond to three different bias-reduced estimators $\hat{m}_{000}^{\text{est}}=\exp \hat{\lambda}^{\text{est}}$. The description of the details on these estimators are beyond the scope of this paper, but they are provided in \cite{Kosmidis2014,Kosmidis2020,Kosmidis2023}. In this paper we limit ourselves to noting that in the DSE and MSE simulation studies presented in this paper we found negligible differences between them, and therefore we denote them as the single estimator $\hat{N}^{\text{CFK,LLM}}$.

In the next section we extend the Chapman-estimator towards multiple sources, which results in a Chapman MSE-estimator that differs from the estimators discussed in this section, both for the saturated and more restricted log-linear models.

\subsection{The Chapman MSE-estimator for saturated models} \label{sec:ChapmanMSE_saturated}
To derive an Chapman MSE estimator we start with the result of \citet[][p. 446]{BishopFienberg1975}, who showed that the MLEs for $m_{abc}$ are equivalent under the assumption that $n_{abc}$ follows either a Poisson or multinomial distribution, provided that $\sum_{abc} m_{abc} + m_{000} = N$. Combining the implications of the Chapman-estimator as discussed in Section \ref{sec:ChapmanBailey} with the MSE models discussed in Section \ref{sec:MSEbasics} under the assumption of a Poisson distribution allows us to derive a bias-corrected MSE estimator in a straightforward way. The Poisson distribution implies that $\Cov\left(n_{abc},n_{\neq abc}\right) = 0$ and $\Cov\left(1/(n_{abc}+1),n_{\neq abc}\right) = \Cov\left(1/n_{abc},n_{\neq abc}\right) = 0$,
when we combine this with Eq. (\ref{eq:correction}) and (\ref{eq:OR_m00...0_exp}) this gives
\begin{align} \label{eq:m000_Chapman_MSE_saturated_general}
m_{00 \ldots 0} &= \frac{\prod{m_{\odd}}}{\prod{m_{\even}}} \approx \prod \mathbf{E} \left[ n_{\odd}\right] \prod \mathbf{E} \left[ \frac{1}{\left(n_{\even}+1\right)}\right] = \mathbf{E} \left[\frac{\prod n_{\odd}}{\prod (n_{\even}+1)}\right], &&
\end{align}
which suggests
\begin{align} 
\hat{m}_{000}^{\text{Chap MSE,SAT}} &= \frac{n_{111}n_{100}n_{010}n_{001}}{(n_{110}+1)(n_{101}+1)(n_{011}+1)} && \label{eq:m000_est_Chapman_MSE_saturated}
\intertext{as a bias-corrected estimator for three sources, and}
\hat{m}_{00 \ldots 0}^{\text{Chap MSE,SAT}} &=  \frac{\prod{n_{\odd}}}{\prod{(n_{\even}}+1)} \label{eq:m000_est_Chapman_MSE_saturated_general}
\end{align}
for any number of sources.

When we compare the Chapman MSE-estimator in Eq. (\ref{eq:m000_est_Chapman_MSE_saturated}) with the RL-estimator in Eq. (\ref{eq:Rivest_other}), it becomes clear that the equality between both estimators in DSE does not hold for MSE. We further note that Chapman MSE estimates can also be obtained with the Poisson regression model as defined in Eq. (\ref{eq:LL_regression}), by using the modified counts $n_{abc}^{\text{Chap MSE,SAT}}$ instead of $\mathbf{n}$.

\subsubsection{Simulation study with saturated models} \label{sec:SimMSE_saturated}
In Section \ref{sec:SimDSE} we have seen that the Chapman- and RL-estimator are equivalent and less biased than the alternative DSE estimators. This equivalence is unlikely to hold in MSE, because they are no longer the same estimators. Here we compare them in a simulation study, together with other bias-reduced MSE estimators. We consider fourteen scenarios. The scenarios in Table \ref{tab:MSE_scenarios} differ with respect to the size of the population $N$, the number of sources $k$ and log-linear model specifications (i.e. different values for $p_{A}$, $p_{B}$, $p_{C}$, $p_{D}$, $\theta^{AB}$, $\theta^{AC}$, $\theta^{AD}$, $\theta^{BC}$ and $\theta^{CD}$, \citealp[see][for further details]{Hammond2023}). The odds-ratios are chosen such that scenario $1-3$ and $13$ concern $\text{LLM}^{\text{sim}}$ = IND, scenario $4-6$ concern $\text{LLM}^{\text{sim}}$ = 1PD, scenario $7-9$ concern $\text{LLM}^{\text{sim}}$ = 2PD, scenario $10-12$ concern $\text{LLM}^{\text{sim}}$ = SAT and finally scenario $14$ concerns $\text{LLM}^{\text{sim}}$ = 4PD (i.e. four pairs of dependent sources), with $\text{LLM}^{\text{sim}}$ the log-linear model used to generate the contingency table. The different parameters are chosen such that the probability of failures in each scenario is small.

\begin{table}[htbp] 
  \caption{MSE simulation scenarios} \label{tab:MSE_scenarios}
\begin{tabular}{ll|rrr@{}lr@{}lr@{}lr@{}lr@{}lr@{}lr@{}lr@{}lr@{}l}
\toprule
$S$ & $\text{LLM}^{\text{sim}}$ & $N$ & $k$ & \multicolumn{2}{c}{$p_{A}$} & \multicolumn{2}{c}{$p_{B}$} & \multicolumn{2}{c}{$p_{C}$} & \multicolumn{2}{c}{} & \multicolumn{2}{c}{$\theta_{AB}$} & \multicolumn{2}{c}{$\theta_{AC}$} & \multicolumn{2}{c}{$\theta_{BC}$} & \multicolumn{2}{c}{} & \\
\midrule
\midrule
$1$  &     & 100    & 3 & 0&.5 & 0&.4  & 0&.3 & & & 1 &   & 1 & & 1 & \\
$2$  & IND & 500    & 3 & 0&.4 & 0&.3  & 0&.2 & & & 1 &   & 1 & & 1 & \\
$3$  &     & 10,000 & 3 & 0&.35 & 0&.3 & 0&.25 & & & 1 &   & 1 & & 1 & \\
\midrule
$4$  &     & 100    & 3 & 0&.5 & 0&.4  & 0&.3 & & & 1&.5 & 1 & & 1 & \\
$5$  & 1PD & 500    & 3 & 0&.4 & 0&.3  & 0&.2 & & & 1&.5 & 1 & & 1 & \\
$6$  &     & 10,000 & 3 & 0&.35 & 0&.3 & 0&.25 & & & 1&.5 & 1 & & 1 & \\
\midrule
$7$  &     & 100    & 3 & 0&.5 & 0&.4  & 0&.3 & & & 1&.5 & 1 & & 0&.5 & \\
$8$  & 2PD & 500    & 3 & 0&.4 & 0&.3  & 0&.2 & & & 1&.5 & 1 & & 0&.5 & \\
$9$  &     & 10,000 & 3 & 0&.35 & 0&.3 & 0&.25 & & & 1&.5 & 1 & & 0&.5 & \\
\midrule
$10$ &     & 100    & 3 & 0&.5 & 0&.4  & 0&.3 & & & 1&.5 &  0&.75 & 0&.5 & \\
$11$ & SAT & 500    & 3 & 0&.4 & 0&.3  & 0&.2 & & & 1&.5 &  0&.75 & 0&.5 & \\
$12$ &     & 10,000 & 3 & 0&.35 & 0&.3 & 0&.25 & & & 1&.5 &  0&.75 & 0&.5 & \\
\midrule
$S$ & LLM & $N$ & $k$ & \multicolumn{2}{c}{$p_{A}$} & \multicolumn{2}{c}{$p_{B}$} & \multicolumn{2}{c}{$p_{C}$} & \multicolumn{2}{c}{$p_{D}$} & \multicolumn{2}{c}{$\theta_{AB}$} & \multicolumn{2}{c}{$\theta_{AD}$} & \multicolumn{2}{c}{$\theta_{BC}$} & \multicolumn{2}{c}{$\theta_{CD}$} \\
\midrule
\midrule
$13$ & IND & 20,000 & 4 & 0&.4  & 0&.35 & 0&.3  & 0&.25 & 1 &   & 1 &    & 1  &   & 1 & \\
$14$ & 4PD & 20,000 & 4 & 0&.4  & 0&.35 & 0&.3  & 0&.25 & 1&.5   &  1&.5 &   0&.75 &  0&.5 \\
\bottomrule
\end{tabular}
\end{table}

\begin{table}[h!]
  \caption{Simulation study with assumed saturated log-linear models, with $20,000$ replications for MSE scenarios $1-14$, for MSE scenario $1-14$ in Table \ref{tab:MSE_scenarios}.} \label{tab:MSE_saturated}
\begin{tabular}{l|rr@{}l|r@{}lr@{}lr@{}lr@{}lr@{}l}
\toprule
$S$ & \multicolumn{1}{r}{$N$} & \multicolumn{2}{r|}{$\bar{n}$} & \multicolumn{2}{c}{$\bar{\hat{N}}^{\text{ML,SAT}}$} & \multicolumn{2}{c}{$\bar{\hat{N}}^{\text{EB,SAT}}$} & \multicolumn{2}{c}{$\bar{\hat{N}}^{\text{CFK,SAT}}$} & \multicolumn{2}{c}{$\bar{\hat{N}}^{\text{RL,SAT}}$} & \multicolumn{2}{c}{$\bar{\hat{N}}^{\text{Chap MSE,SAT}}$} \\
\midrule
1 & 100 &79&.0 & 112&.7$^{\ast\ast\ast\dagger}$ & 112&.4$^{\ast\ast\ast}$ & 110&.9$^{\ast\ast\ast}$ & 103&.3$^{\ast\ast\ast}$ & 100&.1 \\
2 & 500 & 332&.0 & 520&.8$^{\ast\ast\ast}$ & 521&.1$^{\ast\ast\ast}$ & 521&.3$^{\ast\ast\ast}$ & 506&.1$^{\ast\ast\ast}$ & 499&.3 \\
3 & 10,000 &6,587&.7 & 10,015&.5$^{\ast\ast\ast}$ & 10,016&.3$^{\ast\ast\ast}$ & 10,017&.1$^{\ast\ast\ast}$ & 10,004&.1 &9,998&.5 \\
\midrule
4 & 100 &77&.3 & 115&.3$^{\ast\ast\ast\dagger}$ & 114&.4$^{\ast\ast\ast}$ & 111&.7$^{\ast\ast\ast}$ & 103&.4$^{\ast\ast\ast}$ &99&.6$^{\ast}$ \\
5 & 500 & 323&.8 & 525&.2$^{\ast\ast\ast}$ & 524&.4$^{\ast\ast\ast}$ & 523&.8$^{\ast\ast\ast}$ & 508&.3$^{\ast\ast\ast}$ & 500&.7 \\
6 & 10,000 &6,439&.5 & 10,015&.7$^{\ast\ast\ast}$ & 10,016&.0$^{\ast\ast\ast}$ & 10,016&.2$^{\ast\ast\ast}$ & 10,003&.5 &9,997&.4 \\
\midrule
7 & 100 &79&.1 & 121&.5$^{\ast\ast\ast\dagger}$ & 119&.0$^{\ast\ast\ast}$ & 113&.2$^{\ast\ast\ast}$ & 103&.8$^{\ast\ast\ast}$ &99&.5$^{\ast\ast}$ \\
8 & 500 & 330&.3 & 532&.1$^{\ast\ast\ast}$ & 530&.7$^{\ast\ast\ast}$ & 529&.4$^{\ast\ast\ast}$ & 509&.9$^{\ast\ast\ast}$ & 500&.4 \\
9 & 10,000 &6,608&.9 & 10,019&.6$^{\ast\ast\ast}$ & 10,020&.4$^{\ast\ast\ast}$ & 10,021&.1$^{\ast\ast\ast}$ & 10,006&.1$^{\ast}$ &9,999&.3 \\
\midrule
10 & 100 &80&.0 & 117&.8$^{\ast\ast\ast\dagger}$ & 115&.6$^{\ast\ast\ast}$ & 111&.9$^{\ast\ast\ast}$ & 103&.2$^{\ast\ast\ast}$ &99&.4$^{\ast\ast\ast}$ \\
11 & 500 & 334&.1 & 529&.7$^{\ast\ast\ast\dagger}$ & 529&.9$^{\ast\ast\ast}$ & 529&.6$^{\ast\ast\ast}$ & 509&.3$^{\ast\ast\ast}$ & 500&.4 \\
12 & 10,000 &6,690&.3 & 10,019&.6$^{\ast\ast\ast}$ & 10,020&.9$^{\ast\ast\ast}$ & 10,022&.3$^{\ast\ast\ast}$ & 10,006&.4$^{\ast}$ &9,999&.8 \\
\midrule
13 & 20,000 & 15,905&.3 & 20,051&.8$^{\ast\ast\ast}$ & 20,051&.7$^{\ast\ast\ast}$ & 20,051&.6$^{\ast\ast\ast}$ & 20,043&.5$^{\ast\ast\ast}$ & 20,004&.1 \\
14 & 20,000 & 15,834&.5 & 20,049&.4$^{\ast\ast\ast}$ & 20,050&.2$^{\ast\ast\ast}$ & 20,052&.6$^{\ast\ast\ast}$ & 20,039&.9$^{\ast\ast\ast}$ & 19,992&.4 \\
\bottomrule
\end{tabular}
 \begin{tablenotes}
      \small
      \item       
      $\bar{n}$ gives the mean number of observed units $n$ over all replications. The superscripts $^{\ast}$, $^{\ast\ast}$ and $^{\ast\ast\ast}$ indicate that we can reject $\hat{N}^{\text{est}}=N$ with a two-sided t-test with p-values = $0.05, 0.01$ and $0.001$ respectively. A $\dagger$ as superscript indicates that extremely high estimates due to failures were replaced with the highest Chapman MSE estimate in the simulation sample.
    \end{tablenotes}
\end{table}
The estimates presented in Table \ref{tab:MSE_saturated} are based on the saturated model. This means that for all scenarios, except scenario $10-12$, the model is overspecified. Overspecification only affects the variance and not the mean of an estimator, so it does not lead to the introduction of bias, although it may increase the bias when it is present, which is discussed in more detail below Table \ref{tab:MSE_restricted}. In contrast to the DSE simulation study in Section \ref{sec:SimDSE}, it was not possible to exclude failures in all scenarios, in particularly for $N=100$. In those cases the estimates where replaced with the highest value of all Chapman MSE-estimators for that scenario, indicated by a superscript $\dagger$ in the cell. 

The results in Table \ref{tab:MSE_saturated} indicate that, with the saturated model, the Chapman MSE-estimator performs best of the tested estimators, irrespective of the underlying $\text{LLM}^{\text{sim}}$. For $p=0.01$ it gives a mean value that cannot be rejected to be different from $N$ in 13 out of 14 scenarios. Also, in the scenarios where the Chapman MSE-estimator shows some statistically significant bias ($S=4,7$ and $10$), the bias is small in itself and much smaller than in the other estimators. For the IND and 1PD model with large $N$, the bias of the ML, EB and CFK-estimators is equally large. This unexpected indifference might be due to the modification of elements of $n_{abc}$ that are in the numerator of the ML-estimator, which, as we have seen in Section \ref{sec:ChapmanMSE_saturated}, is unnecessary. The RL-estimator performs clearly better than the EB- and CFK-estimator, but still shows some statistically significant bias for most scenarios, especially when $N=100$ or $500$.

The SEs and RMSEs that correspond to each estimator and scenario in Table \ref{tab:MSE_saturated} can be found in Table \ref{tab:MSE_saturated_sd} and \ref{tab:MSE_saturated_rmse} in Appendix \ref{App:sim_saturated}. These tables show that under an assumed saturated model, the Chapman estimator not only outperforms the other estimators in terms of bias, but also in terms of SD and RMSE, in particular for smaller $N$, irrespective of the true model as given in the column $\text{LLM}^{\text{sim}}$.

The estimates in Table \ref{tab:MSE_saturated} are based on the saturated model, but more restricted models such as those in Eq. (\ref{eq:LL3_2PD}), (\ref{eq:LL3_1PD}) and (\ref{eq:LL3_IND}) might be assumed. In the next section we will therefore discuss the Chapman MSE-estimator for restricted models.

\subsection {A generalisation of the Chapman MSE-estimator} \label{sec:MSE_restricted}
The Chapman MSE-estimator for saturated models, as discussed in the previous section, is not necessarily a correct bias-corrected estimator for restricted log-linear models. As an example where the use of the Chapman MSE-estimator for saturated models leads to an incorrect result, consider the 1PD model (\ref{eq:LL3_1PD}) with estimator (\ref{eq:m000_1pairDependence}). When this estimator uses the modified count vector $\mathbf{n}^{\text{Chap MSE, SAT}} = \left(n_{111}, n_{110}+1, n_{101}+1, n_{011}+1, n_{100}, n_{010}, n_{001}\right)$ instead of the observed count vector $\mathbf{n}$, this gives
\begin{align*}
\hat{m}_{000}^{\text{Chap MSE',1PD}} = \frac{n_{001}(n_{++0}+1)}{n_{111}+(n_{101}+1)+(n_{011}+1)}. &&
\end{align*}
We know that this estimator is not correcting for bias correctly, because the ML-estimator for the 1PD model has the same structure as the LP-estimator, namely the product of two Poisson variables in the nominator (i.e. $n_{001}n_{++0}$) and a single Poisson variable in the denominator (i.e. the sum $n_{111}+n_{101}+n_{011}$). Therefore we should use the same bias-correction as used in the Chapman-estimator, namely
\begin{align}
\hat{m}_{000}^{\text{Chap MSE,1PD}} = \frac{n_{001}n_{++0}}{n_{111}+n_{101}+n_{011}+1}. && \label{eq:m000_1pairDependence_bc}
\end{align}
This is the correct bias-corrected estimator for the 1PD model. Similarly, for the 2PD model (\ref{eq:LL3_2PD}) with estimator (\ref{eq:m000_2pairDependence}) we have the bias-corrected estimator
\begin{align}
\hat{m}_{000}^{\text{Chap MSE,2PD}} = \frac{n_{001}n_{100}}{n_{101}+1}. && \label{eq:m000_2pairDependence_bc}
\end{align}
For the independence model for three sources, a direct solution for the estimator does not exist and therefore we cannot use the approach adopted for the 1PD and 2PD model as a general solution. Furthermore, for log-linear models with more sources and more source dependencies, the derivations performed by \cite{BishopFienberg1975} become increasingly complex.

Generally, in order to correct for bias, it appears that we should only know which (functions of) observed counts $n_{abc}$ are in the denominator of $\hat{m}_{000}^{\text{ML,LLM}}$, and subsequently adjust these counts to correct for bias. To identify these (functions of) observed counts $n_{abc}$, we propose to use the Moore-Penrose inverse \cite[MPI,][]{Moore1920,Penrose1955}, that can be used to obtain a \q{best fit} (i.e. least squares) solution (if any exists) for systems of linear equations. 

\subsubsection {Bias reduction by using the Moore-Penrose inverse}\label{sec:OLS_recipe}
We start with the log-linear model in Eq. (\ref{eq:LL_regression}), $\log \mathbb{E} \left[\mathbf{n}|\mathbf{X} \right] = {\mathbf{X}\boldsymbol{\lambda}}$, which is a system of linear equations. A solution for $\boldsymbol{\lambda}$ can be found with the help of the MPI that we write as $\mathbf{Z} = \left(\mathbf{X}^{\top} \mathbf{X}\right)^{-1}\mathbf{X}^{\top}$, i.e.:
\begin{align} \label{eq:MPI}
\boldsymbol{\lambda}^{\text{MPI}} = \mathbf{Z}\log \left[\mathbf{n}|\mathbf{X} \right] = \mathbf{Z}\log \mathbf{m}. &&
\end{align} 
For two sources this gives $\mathbf{m}=\left(m_{11},m_{10},m_{01}\right)^{\top}$, $\boldsymbol{\lambda}^{\text{MPI}}=\left(\lambda^{\text{MPI}},\lambda_{a}^{A,\text{MPI}},\lambda_{b}^{B,\text{MPI}}\right)^{\top}$, $\mathbf{X}=X_{ab}=\left(\begin{array}{ccc} 1 & 1 & 1 \\ 1 & 1 & 0 \\ 1 & 0 & 1 \end{array} \right)$ and $\mathbf{Z} = \mathbf{Z}_{ab} = \left(\begin{array}{r@{}lr@{}lr@{}l} - & 1 & & 1 & & 1 \\ & 1 & & 0 & -& 1 \\ & 1 & -& 1 & & 0 \end{array} \right)$.
We are interested in a solution for $m_{00}$, which is $m_{00} = \exp {\lambda^{\text{MPI}}}$, and because $\lambda^{\text{MPI}}$ depends only on the first row of $\mathbf{Z}_{ab}$, only this row is relevant for our purpose. We write this row as the vector $\mathbf{z} = \left(z_{11},z_{10},z_{01}\right)^{\top}=\left(-1,1,1\right)^{\top}$. Thus (\ref{eq:MPI}) allows us to write $m_{00}$ as a function of $z_{ab}$ and $m_{ab}$, i.e.
\begin{align*}
m_{00} = \exp \lambda^{\text{MPI}} = \exp \sum_{ab} z_{ab} \log m_{ab} = \prod_{ab} (m_{ab})^{z_{ab}} = \frac{m_{10}m_{01}}{m_{11}}, &&
\end{align*}
which corresponds to Eq. (\ref{eq:OR_m}) that led to the LP-estimator, so for two sources $\lambda^{\text{MPI}}=\lambda^{\text{ML}}$. However, for our purpose a more important point is that the first element $z_{11}$ in $\mathbf{z}$ has a negative sign, which indicates that $m_{11}$ is in the denominator of the expression for $m_{00}$. We have seen in Eq. (\ref{eq:m000_Chapman_MSE_saturated_general}), that in order to correct for bias it is important to identify which elements of $\mathbf{n}$ are in the denominator.

This relation between $z_{ab}$ and the LP-estimator also holds for $z_{abc}$ and the SAT, 2PD and 1PD ML-estimators, as defined in Eq. (\ref{eq:m000_OR_SAT}), (\ref{eq:m000_2pairDependence}) and (\ref{eq:m000_1pairDependence}). This can be seen when we specify
\begin{align*}
m_{000}^{\text{MPI,LLM}} = \prod_{abc} \left(m_{abc}\right)^{z_{abc}^{\text{LLM}}}, &&
\end{align*}
where $z_{abc}^{\text{LLM}}$ depends on the design matrices for restricted log-linear models $X_{abc}^{\text{LLM}}$ as defined below Eq. (\ref{eq:LL3_IND}). For the models SAT, 2PD, 1PD and IND, the vector $\mathbf{z}^{\text{LLM}}$ is given in Table \ref{tab:z_abc}a.

\begin{table}[htbp]
\caption{The value of $\mathbf{z}^{\text{LLM}}$ and $\mathbf{z}_{<0}^{\text{LLM}}$ for each LLM.
\label{tab:z_abc}}
\begin{tabular}{ll}
Table \ref{tab:z_abc}a & Table \ref{tab:z_abc}b \\  
\begin{minipage}{.5\linewidth}
        \begin{tabular}{l|r@{}lr@{}lr@{}lr@{}l}
\toprule
$\mathbf{m}$ & \multicolumn{2}{c}{$\mathbf{z}^{\text{SAT}}$} & \multicolumn{2}{c}{$\mathbf{z}^{\text{2PD}}$} & \multicolumn{2}{c}{$\mathbf{z}^{\text{1PD}}$} & \multicolumn{2}{c}{$\mathbf{z}^{\text{IND}}$} \\
\midrule
$m_{111}$ & &1 & &0 & -&1/3 & -&1/2 \\
$m_{110}$ & -&1 & &0 & &1/3 & &0  \\
$m_{101}$ & -&1 & -&1 & -&1/3 & &0  \\
$m_{011}$ & -&1 & &0 & -&1/3 & &0  \\
$m_{100}$ & &1 & &1 & &1/3 & &1/2 \\
$m_{010}$ & &1 & &0 & &1/3 & &1/2  \\
$m_{001}$ & &1 & &1 & &1 & &1/2 \\
\bottomrule
        \end{tabular}
    \end{minipage} &

    \begin{minipage}{.5\linewidth}
        \begin{tabular}{l|r@{}lr@{}lr@{}lr@{}l}
\toprule
$\mathbf{m}$ & \multicolumn{2}{c}{$\mathbf{z}_{<0}^{\text{SAT}}$} & \multicolumn{2}{c}{$\mathbf{z}_{<0}^{\text{2PD}}$} & \multicolumn{2}{c}{$\mathbf{z}_{<0}^{\text{1PD}}$} & \multicolumn{2}{c}{$\mathbf{z}_{<0}^{\text{IND}}$} \\
\midrule
$m_{111}$ & &0 & &0 & -&1/3 & -&1/2 \\
$m_{110}$ & -&1 & &0 & &0 & &0  \\
$m_{101}$ & -&1 & -&1 & -&1/3 & &0  \\
$m_{011}$ & -&1 & &0 & -&1/3 & &0  \\
$m_{100}$ & &0 & &0 & &0 & &0 \\
$m_{010}$ & &0 & &0 & &0 & &0  \\
$m_{001}$ & &0 & &0 & &0 & &0 \\
\bottomrule
        \end{tabular}
    \end{minipage} 
\end{tabular}
\end{table}
Table \ref{tab:z_abc}a shows the positive and negative signs in the elements $\mathbf{z}^{\text{LLM}}$ that correspond to the counts $n_{abc}$ in the numerator and denominator in the SAT, 2PD and 1PD ML-estimators in Eq. (\ref{eq:m000_OR_SAT}), (\ref{eq:m000_2pairDependence}) and (\ref{eq:m000_1pairDependence}).
It is useful to define $\mathbf{z}_{<0}^{\text{LLM}}$, which is a vector equal to $z_{abc}^{\text{LLM}}$ for $z_{abc}^{\text{LLM}}<0$, and zero otherwise. $\mathbf{z}_{<0}^{\text{LLM}}$ is shown in Table \ref{tab:z_abc}b for the SAT, 2PD, 1PD and IND model.

For the 2PD and 1PD model the relation between the MPI expression for $m_{000}$ and the ML-estimator is more intricate. For the 1PD model the MPI expression for $m_{000}$ is
\begin{align*}
m_{000}^{\text{MPI,1PD}} &= \frac{m_{001}\left(m_{110}m_{100}m_{010}\right)^{\frac{1}{3}}}{\left(m_{111}m_{101}m_{011}\right)^{\frac{1}{3}}} &&
\intertext{and the ML-estimator in Eq. (\ref{eq:m000_1pairDependence}) can also be written as}
\hat{m}_{000}^{\text{ML,1PD}} &= \frac{n_{001}\left(m_{110}+m_{100}+m_{010}\right)/3}{\left(m_{111}+m_{101}+m_{011}\right)/3}.
\end{align*}
The MPI expression for $m_{000}$ is a fraction with geometric means of sets of $m_{abc}$, both in the numerator and the denominator, while the ML-estimator is a corresponding fraction of arithmetic means of $n_{abc}$. The same relation can be shown for the SAT and 2PD model. Because a sum of Poisson variables is itself a Poisson variable, and Eq. (\ref{eq:correction}) shows that we should add $1$ to a Poisson variable in the denominator, where $\mathbf{z}_{<0}^{\text{LLM}}$ provides a distribution of this $+1$. To illustrate this, in the bias-corrected estimator for the 1PD model in Eq. (\ref{eq:m000_1pairDependence_bc}), $1$ is added to the sum of the three Poisson variables $n_{111}$, $n_{101}$ and $n_{011}$. The same result is obtained by subtracting $-1/3$ from each of them, which corresponds to subtracting $\mathbf{z}_{<0}^{\text{LLM}}$ from $\mathbf{n}$. Thus we have a simple formula that can be used to obtain the Chapman-estimator in Eq. (\ref{eq:Chap}) and the bias-corrected estimators in Eq. (\ref{eq:m000_OR_SAT}), (\ref{eq:m000_1pairDependence_bc}) and (\ref{eq:m000_2pairDependence_bc}), i.e.:
\begin{align}
\mathbf{n}^{\text{Chap MSE, LLM}} = \mathbf{n}-\mathbf{z}_{<0}^{\text{LLM}} && \label{eq:n_Chapman}
\end{align}
For the IND model Eq. (\ref{eq:n_Chapman}) implies we should replace $n_{111}$ with $n_{111}+1/2$ to obtain a bias-corrected estimator. We cannot compare this result with a closed form expression of the ML-estimator for the IND model, but an intuitive explanation for this adjustment is that if in the denominator there is a Poisson variable multiplied by a $1/2$ as is suggested by the MPI expression, we should add $1$ multiplied by a $1/2$ to correct for bias as well.

Concluding, in Eq. (\ref{eq:n_Chapman}) we propose an adjustment that is based on the MPI and can be used for any log-linear model with any number of sources. We have shown for some examples (i.e., for two sources, and for three sources for the models SAT, 2PD and 1PD) that this adjustment works in these instances. The adjustment also works for the saturated model for any number of sources. We provide no proof for other models, such as the model IND for three sources for which no closed form solutions of ML-estimators exist, or restricted models for four or more sources. In the simulation study in the next section we show that also for these models our procedure reduces the bias to a large extent. Finally, we note that the Chapman MSE adjustment of $n_{abc}$ depends on both the log-linear model and the exact inclusion pattern $abc$, which is more extensive than the information other estimators use.

\subsubsection{Multiple-systems estimation simulation study with restricted models}\label{sec:SimMSE_restricted}
In Table \ref{tab:MSE_restricted} below we show the result of a simulation study in which we test the Chapman MSE estimators under the different scenarios presented in Table \ref{tab:MSE_scenarios}, and compare them with the other estimators described in Section \ref{sec:BiasReducedMSE}. The number of replications is increased from $20,000$ to $60,000$, because in comparison to Table \ref{tab:MSE_saturated}, the estimates are much more accurate because they are based on the same log-linear model that underlies the generation of the contingency table. This is indicated by the $\text{LLM}^{\text{sim}}$ in the subscript. Scenarios $10 - 12$ are removed because they represent scenarios in which the saturated model is the true model, and therefore the results of these scenarios are already provided in Table \ref{tab:MSE_saturated}. The increase of accuracy implies that, compared to the simulation study presented in Section \ref{sec:SimMSE_saturated}, more replications are required to statistically reject unbiasedness in estimators with a t-test.

\begin{table}[htbp]
\renewcommand{\thetable}{\arabic{table}}
  \caption{Simulation study with correctly specified log-linear models, with $60,000$ replications, for MSE scenario $1-9$, $13$ and $14$ in Table \ref{tab:MSE_scenarios}.} \label{tab:MSE_restricted}
\begin{tabular}{r|rr@{}l|r@{}lr@{}lr@{}lr@{}lr@{}l}
\toprule
$S$ & \multicolumn{1}{c}{$N$} & \multicolumn{2}{c|}{$\bar{n}$} & \multicolumn{2}{c}{$\bar{\hat{N}}^{\text{ML,LLM$^{\text{sim}}$}}$} & \multicolumn{2}{c}{$\bar{\hat{N}}^{\text{EB,LLM$^{\text{sim}}$}}$} & \multicolumn{2}{c}{$\bar{\hat{N}}^{\text{CFK,LLM$^{\text{sim}}$}}$} & \multicolumn{2}{c}{$\bar{\hat{N}}^{\text{RL,LLM$^{\text{sim}}$}}$} & \multicolumn{2}{c}{$\bar{\hat{N}}^{\text{Chap MSE,LLM$^{\text{sim}}$}}$} \\
\midrule
1 & 100 &79&.0 & 100&.5$^{\ast\ast\ast}$ & 100&.7$^{\ast\ast\ast}$ & 100&.9$^{\ast\ast\ast}$ & 100&.7$^{\ast\ast\ast}$ &99&.9$^{\ast\ast\ast}$ \\
2 & 500 & 332&.0 & 501&.5$^{\ast\ast\ast}$ & 501&.1$^{\ast\ast\ast}$ & 501&.9$^{\ast\ast\ast}$ & 501&.3$^{\ast\ast\ast}$ & 499&.9 \\
3 & 10,000 &6,587&.3 & 10,001&.0$^{\ast}$ & 10,000&.6 & 10,001&.4$^{\ast\ast}$ & 10,000&.8 &9,999&.4 \\
\midrule
4 & 100 &77&.3 & 101&.2$^{\ast\ast\ast}$ & 101&.3$^{\ast\ast\ast}$ & 102&.1$^{\ast\ast\ast}$ &99&.9 & 100&.0 \\
5 & 500 & 323&.8 & 503&.5$^{\ast\ast\ast}$ & 502&.6$^{\ast\ast\ast}$ & 504&.2$^{\ast\ast\ast}$ & 499&.6$^{\ast}$ & 500&.2 \\
6 & 10,000 &6,439&.6 & 10,003&.1$^{\ast\ast\ast}$ & 10,002&.5$^{\ast\ast\ast}$ & 10,003&.7$^{\ast\ast\ast}$ & 10,000&.1 & 10,000&.4 \\
\midrule
7 & 100 &79&.1 & 102&.7$^{\ast\ast\ast\dagger}$ & 102&.8$^{\ast\ast\ast}$ & 102&.8$^{\ast\ast\ast}$ & 100&.8$^{\ast\ast\ast}$ &99&.9 \\
8 & 500 & 330&.3 & 505&.9$^{\ast\ast\ast}$ & 505&.6$^{\ast\ast\ast}$ & 505&.3$^{\ast\ast\ast}$ & 501&.7$^{\ast\ast\ast}$ & 499&.8 \\
9 & 10,000 &6,608&.9 & 10,005&.0$^{\ast\ast\ast}$ & 10,004&.8$^{\ast\ast\ast}$ & 10,004&.7$^{\ast\ast\ast}$ & 10,001&.5 &9,999&.7 \\
\midrule
13 & 20,000 & 15,320&.0 & 20,000&.3 & 19,999&.9 & 20,000&.7 & 20,000&.3 & 19,999&.4 \\
14 & 20,000 & 15,166&.8 & 20,002&.4$^{\ast\ast\ast}$ & 20,002&.3$^{\ast\ast}$ & 20,002&.4$^{\ast\ast\ast}$ & 20,001&.4$^{\ast}$ & 20,000&.5 \\
\bottomrule
\end{tabular}
 \begin{tablenotes}
      \small
      \item[]      
      $\bar{n}$ gives the mean number of observed units $n$ over all replications. $\text{LLM}^{\text{sim}}$ in the superscript  indicates that the estimates are obtained under the correctly specified log-linear model, as given in the \q{$\text{LLM}^{\text{sim}}$} column in Table \ref{tab:MSE_scenarios}. The superscripts $^{\ast}$, $^{\ast\ast}$ and $^{\ast\ast\ast}$ indicate that we can reject $\hat{N}^{\text{est}}=N$ with a two-sided t-test for p-values = $0.05, 0.01$ and $0.001$ respectively. A $\dagger$ as superscript indicates that extremely high estimates due to failures were replaced with the highest Chapman MSE estimate in the simulation sample.
    \end{tablenotes}
\end{table}

Table \ref{tab:MSE_restricted} shows that if the correct log-linear model is used to estimate $N$, all estimators have less bias than under the saturated model as presented in Table \ref{tab:MSE_saturated}. This is unsurprising, because for the presented scenarios the model that is used for estimation is less overspecified than the saturated model that was used for Table \ref{tab:MSE_saturated}. For $S=13$, the bias is statistically insignificant in all estimators. The size of the bias in the ML-, EB- and CFK-estimator is comparable and small, and the RL-estimator has less bias than these three estimators. The Chapman MSE-estimator is the only estimator for which unbiasedness cannot be statistically rejected in all except for one scenario. Only in $S=1$ there is some statistically significant bias, but this bias is small in itself and less than the bias in the other estimators. To conclude, also when the correct (restricted) model is used to estimate $N$, the Chapman MSE-estimators outperforms the other estimators, although for simpler models the impact becomes less substantial because the bias is smaller in the first place.

The substantial difference in the magnitude of the bias shown in Tables \ref{tab:MSE_saturated} and \ref{tab:MSE_restricted}, is caused by the inflationary effect of variance on positive bias in log-linear models. Less restricted models have larger variance \citep[][p. 242]{BishopFienberg1975} and its inflationary effect on the bias can be seen when the bias is written as
$\bar{\hat{m}}_{000}^{\text{est}} - m_{000} = \overline{\exp \hat{\lambda}^{\text{est}}} - \exp \lambda$, 
where given some positive bias in $\hat{\lambda}^{\text{est}}$, a larger variance in $\hat{\lambda}^{\text{est}}$ leads to a further increase of $\overline{\exp \hat{\lambda}^{\text{est}}}$ and therefore the bias.

Particularly interesting are scenarios for which we were not able to prove mathematically that our MPI procedure is correct, i.e. $1$, $2$, $3$, $13$ and $14$, as these are scenarios in which the independence model and/or four sources are used. In these scenarios the Chapman MSE-estimator provides estimates with clearly less bias than the bias in the other estimators, and also little bias in general.
This is support for the approach that we adopted for the development of the bias-reduced estimators based on Eq. (\ref{eq:n_Chapman}).

The SDs and RMSEs that correspond to each estimator and scenario in Table \ref{tab:MSE_restricted} can be found in Table \ref{tab:MSE_restricted_sd} and  \ref{tab:MSE_restricted_rmse} in Appendix \ref{App:sim_restricted}. These tables show that, just like the bias, also the SD and RMSE are smaller when the correctly specified log-linear model is used for estimation. Also in this case, the RL- and Chapman MSE-estimator outperform the other estimators, especially for models with more parameters. Similar to bias, for the 2PD and SAT model (for SAT see Table \ref{tab:MSE_saturated}, \ref{tab:MSE_saturated_sd} and \ref{tab:MSE_saturated_rmse}) we see that the Chapman MSE-estimator outperforms the RL-estimator in terms of SD and RMSE. This shows that the correction for bias becomes more important when the estimated model has more parameters.

Finally, we note that the Chapman MSE-estimator suffers less from adding irrelevant variables to the model. To illustrate this we consider scenario $S=7$ from Table \ref{tab:MSE_scenarios}, for which the saturated model contains the irrelevant parameter $\lambda_{ac}^{AC}$. When we consider the SDs of the Chapman MSE-estimator for this scenario, as shown in Table \ref{tab:MSE_saturated_sd} and \ref{tab:MSE_restricted_sd}, we see that it approximately doubles from $12.3$ to $24.7$. For the other estimators adding the irrelevant parameter $\lambda_{ac}^{AC}$ has a larger impact on the SD, as it increases approximately three to six times while starting from approximately the same level. The same relation holds for the RMSE.

\section{Example: Number of homeless people in the Netherlands}  \label{sec:Homeless}
A population size estimate of the homeless people in the Netherlands is published annually by Statistics Netherlands. This estimate is an ML estimate that is based on a MSE model that is discussed in detail in \cite{Coumans2017}. The estimate is based on a log-linear model that contains three sources and several (categorical) covariates, such as gender ($g$, $2$ categories), age ($a$, $3$ categories), place of living, in- or outside one of the big four Dutch cities ($p$, $2$ categories) and region of origin ($o$, $3$ categories). Together there are $36$ subgroups that have observed frequencies denoted as $n_{gapo}$ and an observed frequency with a specific inclusion pattern denoted as $n_{abc,gapo}$. Which sources, covariates and interactions between them are included in the log-linear model, is the result of an Akaike information criterion (AIC) based model selection procedure that is explained in 
\cite{Coumans2017}. Recent work by \cite{Silverman2020} suggests that other model selection approaches based on Bayesian approaches could lead to more robust and stable results, but this is beyond the scope of this paper.

In this practical example, for the years 2009 - 2018, 2020 and 2021, we replicate the model selection and estimation procedure as explained in \cite{Coumans2017}. Data for 2019 is unavailable. This gives a series of annual ML estimates for the population size of homeless people in The Netherlands. For each year, the log-linear model that was used to calculate the ML estimate is also used to calculate the corresponding Chapman MSE estimate. This allows us to calculate the difference between the ML and Chapman MSE estimates, all other factors held constant, in a practical example.

\begin{figure}[htbp]
\caption{Total number of homeless people, homeless men and homeless women in the Netherlands over the period 2009-2018 and 2020-2021.} \label{fig:Homeless}
1a: All homeless people \\
\begin{subfigure}
     \centering \includegraphics[scale=0.62, trim = {0 1.6cm 0 1.5cm}]{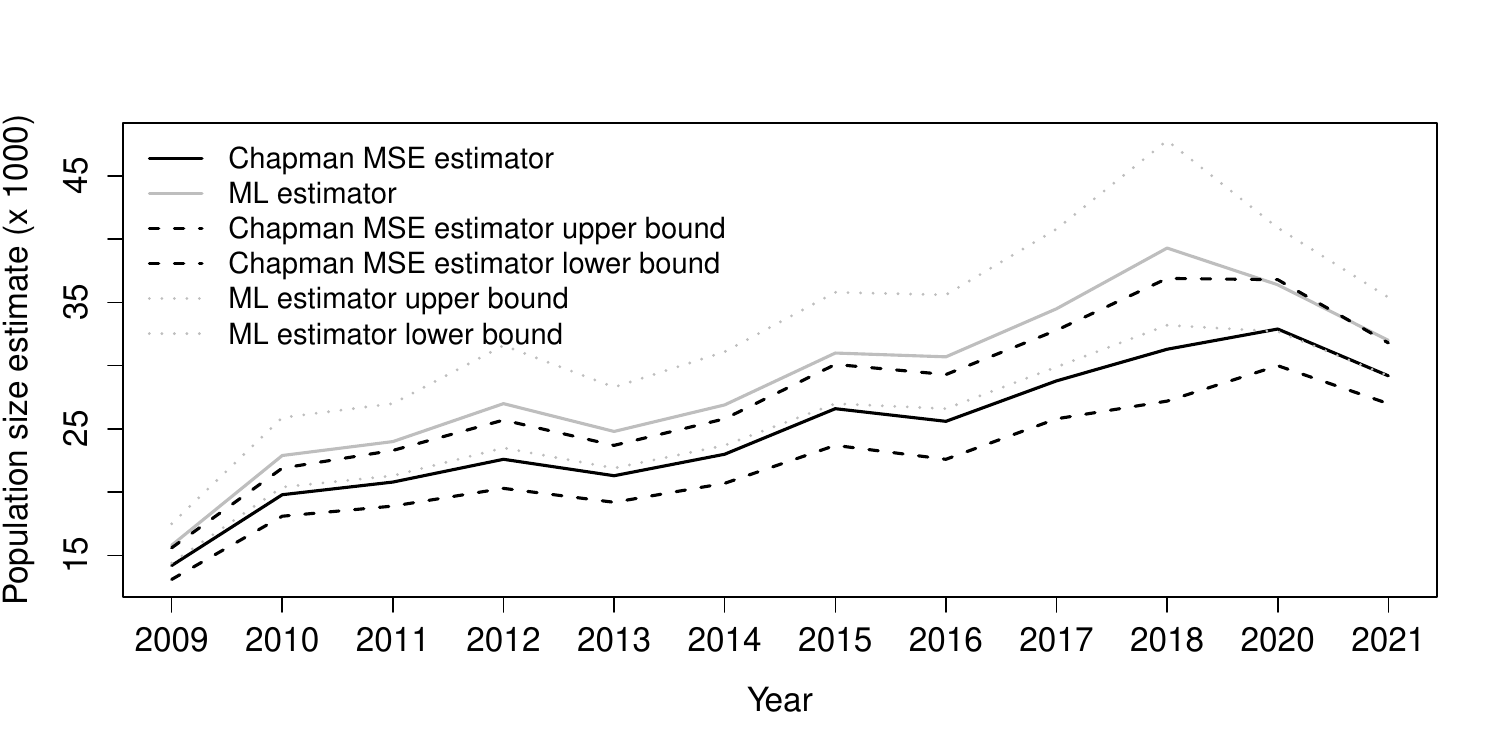}
    \end{subfigure} \\
1b: Homeless men \\
\begin{subfigure}
     \centering \includegraphics[scale=0.62, trim = {0 1.6cm 0 1.5cm}]{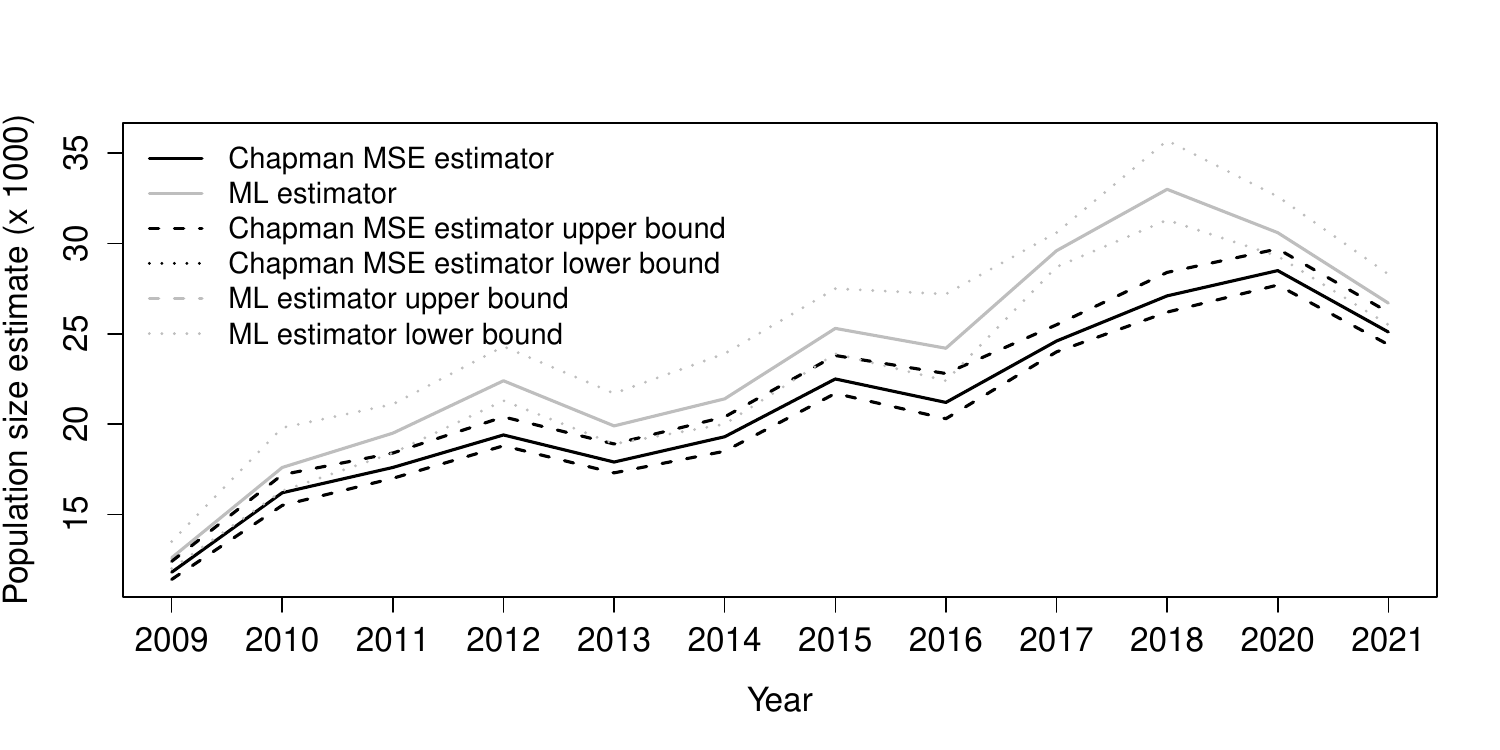}
    \end{subfigure} \\
1c: Homeless women \\
\begin{subfigure}
     \centering \includegraphics[scale=0.62, trim = {0 1.6cm 0 1.5cm}]{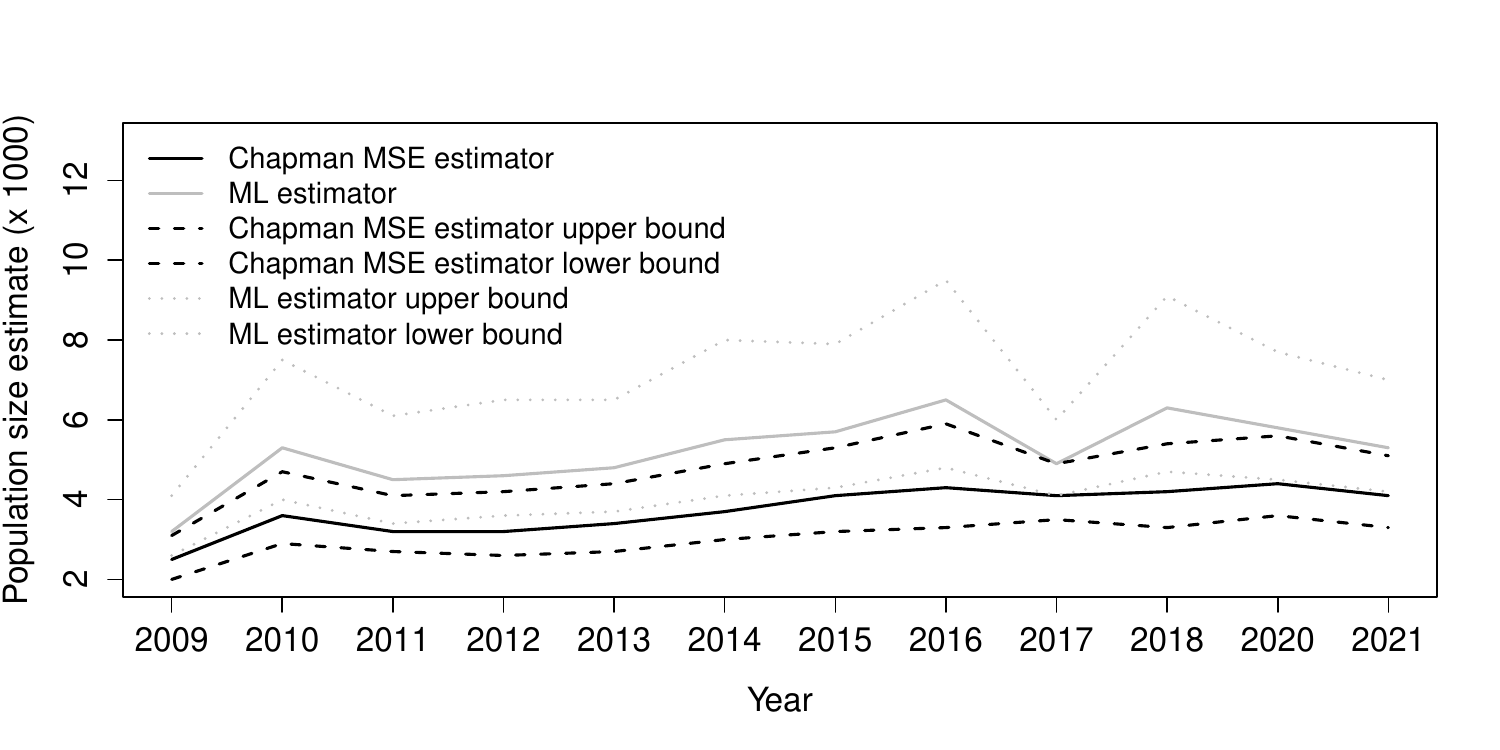}
    \end{subfigure}
  \end{figure}

In Figures \ref{fig:Homeless}a-c below we show, respectively, the original ML estimates and the Chapman MSE estimates of the total number of homeless people, the total number of homeless men and the total number of homeless women, including their two-sided 95\% confidence intervals. Note that each figure has its own scale on the y-axis.

\FloatBarrier
Figure \ref{fig:Homeless}a shows ML and Chapman MSE estimates of the total number of homeless people over time, together with their confidence intervals. The ML estimates are between a minimum of $9.5\%$ and a maximum of $25.5\%$ larger than the Chapman MSE-estimator. The confidence interval of the Chapman MSE-estimators is clearly smaller. Figure \ref{fig:Homeless}b and \ref{fig:Homeless}c show that the total annual difference between both estimators, as was observed in Figure \ref{fig:Homeless}a, is not proportionally divided over men and women. In fact, the Chapman MSE-estimator has, relatively, a much larger impact on the estimate of the number of homeless women, which is the smaller group. For women the difference between the estimates is between a minimum of $19.5\%$ in 2017 and a maximum of $51.2\%$ in 2018.

In this practical application the impact of using the Chapman MSE-estimator instead of the ML-estimator is larger than the impact we have seen in the simulation studies. The reason for this difference is twofold. First, the scenarios in the simulation studies were set such that the probability of estimation failures was very small, which led to a mean coverage (i.e. $\bar{n}/N$) that was large compared to the coverage in our example of homeless people. Second, the MSE model to estimate the number of homeless people involves the use of (categorical) covariates to control for heterogeneity in inclusion probabilities. Because for some homeless people their background characteristics are missing, the estimation procedure uses an expectation–maximization (EM) algorithm to impute missing data \citep[see][for further details]{Coumans2017}, which for some inclusion patterns may lead to observed frequencies between zero and one. To see why this is important we zoom in on the underlying subgroup estimates for men and women in the year 2021 presented in Table \ref{tab:Homeless} below.

Table \ref{tab:Homeless} presents $18$ subgroups indicated by $G_{apo}$ for both men and women. For each subgroup we show both the total observed count $n_{gapo}$ and the observed count $n_{101,gapo}$ for inclusion pattern $101$. This specific inclusion pattern is shown because the selected log-linear model is a 2-pair dependence model, for which Table \ref{tab:z_abc} tells us that $n_{101}^{\text{Chap}}=n_{101}+1$ is the only adjusted observed frequency, while the other elements in $n_{abc}^{\text{Chap}}$ are equal to $n_{abc}$. The difference between $n_{101,gapo}$ and $n_{101,gapo}^{\text{Chap}}$ should therefore explain the difference between $N_{gapo}^{\text{ML}}$ and $N_{gapo}^{\text{Chap}}$.  This difference is shown in the columns $\Delta_{Mapo}^{\text{Chap-ML}}$ and $\Delta_{Wapo}^{\text{Chap-ML}}$.

\begin{table}[htbp]
  \caption{Estimated number of homeless people in The Netherlands in 2021, separated by men and women and $18$ subgroups based on age, living in- or outside one of the four big Dutch cities and country of origin.\label{tab:Homeless}}
\scalebox{0.85}{
\begin{tabular}{r|rr@{}lrrr|rr@{}lrrr}
\toprule
& \multicolumn{6}{c|}{\textbf{Men}} & \multicolumn{6}{c}{\textbf{Women}} \\
\midrule
$G_{apo}$ & $n_{Mapo}$ & \multicolumn{2}{c}{$n_{101,Mapo}$} & $\hat{N}_{Mapo}^{\text{ML}}$ & $\hat{N}_{Mapo}^{\text{Chap}}$ & $\Delta_{Mapo}^{\text{Chap-ML}}$ & $n_{Wapo}$ & \multicolumn{2}{c}{$n_{101,Wapo}$} & $\hat{N}_{Wapo}^{\text{ML}}$ &  $\hat{N}_{Wapo}^{\text{Chap}}$ & $\Delta_{Wapo}^{\text{Chap-ML}}$ \\
\midrule
1 & 1,956 & 134&.07 &4,279 &4,263 & -16 &388 &8&.10 &787 &678 & -109 \\
2 & 1,283 &45&.78 &4,687 &4,464 &-223 &211 &4&.03 &993 &750 & -243 \\
3 & 1,130 &37&.41 &4,458 &4,304 &-154 &164 &2&.56 &760 &582 & -178 \\
4 &516 &17&.62 &1,006 & 978 & -28 & 97 &1&.52 &170 &147 & -23 \\
5 &496 & 9&.56 &2,241 &2,065 &-176 & 76 &0&.90 &333 &245 & -88 \\
6 &491 &41&.02 &1,316 &1,278 & -38 &123 &3&.65 &325 &264 & -61 \\
7 &436 &36&.36 &1,072 &1,055 & -17 &102 &2&.82 &243 &202 & -41 \\
8 &350 &12&.83 &1,388 &1,302 & -86 & 52 &1&.11 &279 &204 & -75 \\
9 &319 &11&.04 &1,224 &1,133 & -91 & 57 &1&.24 &314 &226 & -88 \\
10 &241 & 7&.72 & 555 & 533 & -22 & 45 &0&.66 & 92 & 77 & -15 \\
11 &237 & 6&.07 &1,222 & 989 &-233 & 47 &0&.63 &311 &198 & -113 \\
12 &224 & 4&.84 & 952 & 890 & -62 & 35 &0&.46 &142 &107 & -35 \\
13 &201 &11&.23 & 685 & 586 & -99 & 55 &1&.02 &181 &130 & -51 \\
14 &106 & 2&.71 & 329 & 274 & -55 & 25 &0&.29 & 66 & 48 & -18 \\
15 & 95 & 7&.82 & 287 & 275 & -12 & 28 &0&.90 & 89 & 70 & -19 \\
16 & 91 & 1&.44 & 561 & 435 &-126 & 17 &0&.17 &104 & 65 & -39 \\
17 & 46 & 1&.15 & 252 & 194 & -58 &9 &0&.14 & 72 & 45 & -27 \\
18 & 35 & 1&.90 & 150 & 120 & -30 & 11 &0&.24 & 50 & 34 & -16 \\
\midrule
Total & 8,253 & 390&.57 & 26,664 & 25,138 & -1,526 & 1,542 & 30&.44 & 5,311 & 4,072 & -1,239 \\
\bottomrule
\end{tabular}
}
\end{table}

When we compare the columns $\Delta_{Mapo}^{\text{Chap-ML}}$ and $\Delta_{Wapo}^{\text{Chap-ML}}$ in Table \ref{tab:Homeless}, we see that despite the fact that observed counts of men are larger than those of women, differences in counts of subgroups of men and women are very similar. This can be explained by the smaller observed frequencies for women with inclusion pattern $101$, that are sometimes even between zero and one, as can be seen in the columns of $n_{101,Mapo}$ and $n_{101,Wapo}$. Adding $1$ to such a small number has a relatively large impact on the population size estimate.

Finally, we note that the Chapman MSE estimates follow a similar trend to the ML-estimates, which is relevant in practice. Only for the period $2018 - 2020$, where the ML-estimate is a decrease while the Chapman MSE-estimate is an increase. This might be due to the large ML-estimate in $2018$. Furthermore, the estimates and conclusions presented in this section should be treated with some care because for the log-linear model that was used, a regularity condition such as the one for DSE given by Chapman in Eq (\ref{eq:regularity_DSE}) may play a role. The fact that such a regularity condition for MSE is unknown is unfortunate, because some of the subgroups are quite small and so the risk of not meeting a potential regularity condition is not unrealistic. The data on homeless people in The Netherlands that were used for this section is not publicly available due to legal restrictions.

\section{Discussion} \label{sec:Discussion}
In this paper we have derived the Chapman MSE-estimator and we have shown that, in terms of mean-bias correction, it outperforms a set of other bias-reduced MSE estimators known in literature. We showed both mathematically and in a simulation study that mean-bias correction in DSE is best achieved by means of the DSE estimator proposed by \cite{Chapman1951} and later by \cite{Rivest2001}. Furthermore we showed how the Chapman-estimator can be derived in a different way than \citeauthor{Chapman1951} did. This derivation was extended towards multiple sources, which led to the Chapman MSE-estimator for saturated models. We developed the Chapman MSE-estimator such that it can be applied under both a saturated and restricted model. This generalisation was achieved by using the MPI and for a small set of different restricted models it was proven mathematically that this approach leads to bias-corrected estimators. We used a simulation study to investigate bias in a larger set of restricted models and we found that also for these models the Chapman MSE-estimator shows little or no bias.

The mathematical derivations and simulation studies in this paper show that for any restricted model with three sources or a saturated model with any number of sources, the Chapman MSE-estimator is a bias-corrected estimator. We suspect that this result can be generalised towards any restricted model with any number of sources, although we did not provide a mathematical proofs. We think that further research that proves, or disproves, our suspicion would be valuable.

The simulation studies also show that the Chapman MSE-estimator outperforms other estimators in terms of a smaller size of bias and SD, and thus RMSE, in particular when the estimated log-linear model has more interaction parameters. This advantage is important because in practice the model that is used is usually the result of some model selection procedure, which does not guarantee the selection of the correct model. When such a selection procedure selects a model with irrelevant parameters, this increases the variance of the population size estimate. This increase is less for the Chapman MSE-estimator than for the other estimators considered. 

In Section \ref{sec:Homeless} we applied the Chapman MSE-estimator to estimate the number of homeless people in The Netherlands for a series of years and compared these estimates with the ML estimates. For each year both estimates are based on the same log-linear model as discussed in \cite{Coumans2017}. This comparison showed that the impact of bias-correction can be substantial, e.g., in our example the use of the Chapman MSE-estimator led to a Chapman MSE estimate that was between 9.3\% and 25.4\% lower for the total number of homeless people in The Netherlands, as compared to the corresponding ML-estimator. This relative difference became even larger, going up to 51\%, when we zoomed in on the subgroup of women. 

The simulation studies and the example in Section \ref{sec:Homeless} show that the difference between the Chapman MSE- and the standard ML-estimator can be substantial. This raises the question whether finite-sample bias correction should not have a more prominent role in the discussion on the robustness of MSE methodology and the accuracy of MSE estimates, which continues up till today \cite[see e.g.][]{Silverman2020,Binette2022}.

Finally, a topic that received little or no attention in MSE literature, but what would be valuable to investigate, is regularity conditions. Chapman gave a regularity condition for his DSE estimator, but similar regularity conditions for MSE estimators are unknown. This topic is also beyond the scope of this paper but we think that this is an important remaining problem for MSE estimators in general, including the Chapman MSE-estimator.

\subsection*{Software} \label{sec:Sofware}
All simulation studies in this paper are performed in the statistical software program R \citep{R2022}. All estimates are obtained with the {\fontfamily{qcr}\selectfont glm()} function, with {\fontfamily{qcr}\selectfont family = poisson(link = "log")}. Differences between the LP, ML, Chapman, Bailey, EB, RL and Chapman MSE estimates are the sole result of different input vectors $\mathbf{n}^{\text{est}}$. For the IND model the estimation results for the RL-estimator were verified with the function {\fontfamily{qcr}\selectfont closedp.bc()} with {\fontfamily{qcr}\selectfont m = "Mt"} from the R-package {\fontfamily{qcr}\selectfont Rcapture} \citep{Rivest2022}. The Cordeiro-, Firth- and Kosmidis-estimator ($\hat{N}^{\text{CFK}}$) were also calculated with the {\fontfamily{qcr}\selectfont glm()} function, but with the additional settings {\fontfamily{qcr}\selectfont method = "brglmFit"} and {\fontfamily{qcr}\selectfont type = "correction"},  {\fontfamily{qcr}\selectfont type = "AS\_mean"} and {\fontfamily{qcr}\selectfont type = "MPL\_Jeffreys"}, respectively, which are part of the R-package {\fontfamily{qcr}\selectfont brglm2} \citep{Kosmidis2023}. Code for the simulation studies presented in this paper is available at \\
\href{https://github.com/DaanZult/ChapmanMSE/}{https://github.com/DaanZult/ChapmanMSE/}.

\section*{Author contributions statement}
D.Z. derived the Chapman MSE-estimator, did the set up and programming of the simulation studies, and wrote the manuscript. P.H. had the initial idea for the study and edited the manuscript. B.B. edited the manuscript.

\section*{Acknowledgements}
The authors thank Jeroen Pannekoek, Peter-Paul de Wolf, Sander Scholtus and Moniek Coumans from Statistics Netherlands for their detailed comments and suggestions on this paper.

\clearpage
\bibliography{referencesBCinMSE.bib}
\appendix

\newpage%
 \renewcommand{\thesection}{\Alph{section}}

\section{Comparison of Taylor approximation and Stephan's inverse factorial approximation} \label{App:Taylor_vs_IF}

The Taylor expansion that was also used by \cite{Bailey1951} is a widely used approximation approach, but it is not always the most accurate or efficient method to approximate a function. To illustrate that the inverse factorial (IF) expansion \cite[see e.g.][]{Stephan1945} used by \cite{Chapman1951} gives more accurate results for $\mathbb{E} \left[ \frac{1}{n_{11}} \right]$, given the same number of expansion terms, than a Taylor expansion, we provide a straightforward simulation study. With $r$ replications of $n_{11,r} \sim \Poisson(m_{11})$, we can write five-term expansion Taylor and IF approximations for $\mathbb{E} \left[ \frac{1}{n_{11}} \right]$ as
\begin{align*}
\text{Taylor} \rightarrow \mathbb{E} \left[ \frac{1}{n_{11}}           \right] = \mathbb{E} & \left[ \frac{1}{m_{11}} - 
    \frac{(n_{11} - m_{11})}{m^{2}} +
    \frac{(n_{11} - m_{11})^{2}}{m_{11}^{3}} - \right. \\
    & \left. \frac{(m_{11} - n_{11})^{3}}{m_{11}^{4}} +
    \frac{(m_{11} - n_{11})^{4}}{m_{11}^{5}} - \ldots \right] && 
    \intertext{where $m_{11}$ will be estimated by $\hat{m}_{11} = \sum_{r} n_{11,r}/r$, and}
\text{IF} \rightarrow \mathbb{E} \left[ \frac{1}{n_{11}} \right] \approx \sum_{r} & \left(\frac{1}{n_{11,r}+1}\right)/r + 
    \sum_{r} \left(\frac{1}{\left(n_{11,r}+1\right)\left(n_{11,r}+2\right)}\right)/r + \\
    \sum_{r} & \left(\frac{2}{\left(n_{11,r}+1\right)\left(n_{11,r}+2\right)\left(n_{11,r}+3\right)}\right)/r + \\
    \sum_{r} & \left(\frac{6}{\left(n_{11,r}+1\right)\left(n_{11,r}+2\right)\left(n_{11,r}+3\right)\left(n_{11,r}+4\right)}\right)/r + \\
    \sum_{r} & \left(\frac{24}{\left(n_{11,r}+1\right)\left(n_{11,r}+2\right)\left(n_{11,r}+3\right)\left(n_{11,r}+4\right)\left(n_{11,r}+5\right)}\right)/r
\end{align*}
Table \ref{tab:Poisson} shows the results for both approximation methods and their difference $\Delta$ for $m_{11}=20$ and $r = $ one million.
\begin{table}[H]
\setcounter{table}{0}
\renewcommand{\thetable}{A\arabic{table}}
  \caption{Simulated approximations of $\mathbf{E} \left[\frac{1}{n_{11}}\right]$, with $n_{11,r} \sim \Poisson(m_{11}=20)$ and $r =$ one million, which gives $\mathbf{E} \left[\frac{1}{n_{11}}\right] \approx \left(\sum_{r} \frac{1}{n_{11,r}}\right)/r = 0.052805$.\label{tab:Poisson}}
\begin{tabular}{l|cc|cc}
\toprule
\textbf{\# Terms} & \textbf{Taylor} & $\Delta(\mathbf{E} \left[\frac{1}{n_{11}}\right]-\textbf{Taylor})$ & \textbf{IF} & $\Delta(\mathbf{E} \left[\frac{1}{n_{11}}\right]-\textbf{IF})$ \\[0.20cm]  
\midrule
$1$ & 0.050001 & 0.002804 & 0.050006 & 0.002799 \\
$2$ & 0.050001 & 0.002804 & 0.052507 & 0.000298 \\
$3$ & 0.052505 & 0.000299 & 0.052757 & 0.000048 \\
$4$ & 0.052379 & 0.000426 & 0.052794 & 0.000010 \\
$5$ & 0.052763 & 0.000042 & 0.052802 & 0.000003 \\
\bottomrule
\end{tabular}
\end{table}
Table \ref{tab:Poisson} shows that, for $n_{11,r} \sim \Poisson(m_{r}=20)$ and five or less expansion terms, the IF approximation method used by \cite{Chapman1951} gives a more accurate approximation of $\mathbf{E} \left[\frac{1}{n_{11}}\right] \approx \left(\sum_{r} \frac{1}{n_{11,r}}\right)/r = 0.052805$ than the Taylor approximation method.
\\
\\

\section{Second-order Taylor approximation of the Lincoln-Petersen-estimator \label{App:BiasinDSE}}
Here we present an alternative derivation of a bias-reduced LP-estimator. This derivation shows that the Chapman-estimator can be approximated with the well-known Taylor approximation. We write the LP-estimator as a Taylor series approximation. When we start with some function $f(\mathbf{n})$ of the three random variables $n_{11}$,$n_{10}$ and $n_{01}$, and approximate it around $\mathbf{m}$, this gives:

\begin{align*}
f(\mathbf{n}) = f(\mathbf{m}) + (\mathbf{n}-\mathbf{m})^{\top} \nabla f(\mathbf{m}) + \frac{1}{2} (\mathbf{n}-\mathbf{m})^{\top} \nabla\nabla f(\mathbf{m}) (\mathbf{n}-\mathbf{m}) \nonumber + O(||(\mathbf{n}-\mathbf{m})^{\top}||)^{2}) &&
\end{align*}
with
\begin{align*}
\nabla f(\mathbf{m}) &= \left(\begin{array}{c} 
							\frac{\partial f(\mathbf{n})}{\partial n_{11}} \\
							\frac{\partial f(\mathbf{n})}{\partial n_{10}} \\
							\frac{\partial f(\mathbf{n})}{\partial n_{01}}
						 \end{array}\right)_{\mathbf{m}}  &&
\shortintertext{and}
\nabla\nabla f(\mathbf{m}) &= \left(\begin{array}{ccc} 
					\frac{\partial^2 f(\mathbf{n})}{\partial n_{11}^2} &  \frac{\partial^2 f(\mathbf{n})}{\partial n_{11}\partial n_{10}} & \frac{\partial^2 f(\mathbf{n})}{\partial n_{11}\partial n_{01}} \\
					\frac{\partial^2 f(\mathbf{n})}{\partial n_{10}\partial n_{11}} &  \frac{\partial^2 f(\mathbf{n})}{\partial n_{10}^2} & \frac{\partial^2 f(\mathbf{n})}{\partial n_{10}\partial n_{01}}\\
					\frac{\partial^2 f(\mathbf{n})}{\partial n_{01}\partial n_{11}} &  \frac{\partial^2 f(\mathbf{n})}{\partial n_{01}\partial n_{10}} & \frac{\partial^2 f(\mathbf{n})}{\partial n_{01}^2}	\end{array}\right)_{\mathbf{m}}
\end{align*}
Replacing $f(\mathbf{n})$ with $\hat{m}_{00}^{\text{LP}}=\frac{n_{10}n_{01}}{n_{11}}$ gives:
\begin{align*}
\nabla f(\mathbf{n}) &=  \left(\begin{array}{c}
                        -\frac{n_{10}n_{01}}{n_{11}^2} \\
					\frac{n_{10}}{n_{11}} \\
					\frac{n_{01}}{n_{11}}					
				 \end{array}\right)  &&
\shortintertext{and}
\nabla\nabla f(\mathbf{n}) &= \left(\begin{array}{ccc} 
					\frac{2n_{10}n_{01}}{n_{11}^3} & -\frac{n_{01}}{n_{11}^2}  & -\frac{n_{10}}{n_{11}^2} \\
					-\frac{n_{01}}{n_{11}^2} & 0 & \frac{1}{n_{11}}\\
					-\frac{n_{10}}{n_{11}^2} & \frac{1}{n_{11}} & 0
					 \end{array}\right).
\end{align*}

Therefore, because $\mathbb{E} \left[ (\mathbf{n}-\mathbf{m})^{\top} \nabla f(\mathbf{m}) \right] = 0$, we find:
\begin{align} \label{eq:bias_decomposed}
\mathbb{E} \left[\frac{n_{10}n_{01}}{n_{11}} \right] \approx 
& \frac{m_{10}m_{01}}{m_{11}} + && \nonumber \\
& \frac{\Cov\left(n_{10},n_{01}\right)}{m_{11}} -
 \frac{m_{10}\Cov\left(n_{11},n_{01}\right)}{m_{11}^2} 
- \frac{m_{01}\Cov\left(n_{11},n_{10}\right)}{m_{11}^2} + \nonumber \\
& \frac{m_{10}m_{01}\Var\left(n_{11}\right)}{m_{11}^3}.
\end{align}
For the Poisson distribution we have $\Cov\left(n_{ab},n_{\neq ab}\right) = 0$ and $\Var\left(n_{ab}\right) = m_{ab}$, and for the multinomial distribution we have $\Cov\left(n_{ab},n_{\neq ab}\right) = -Np_{ab}p_{\neq ab}$ and $\Var\left(n_{ab}\right) = Np_{ab}(1-p_{ab})$ with $p_{ab} = m_{ab}/N$.
Then, for both $n_{ab} \sim \Poisson(m_{ab})$ and the joint set $(n_{11},n_{10},n_{01},n_{00}) \sim \text{Multinomial}(m_{11},m_{10},m_{01},m_{00})$, Eq. (\ref{eq:bias_decomposed}) reduces to:
\begin{align} \label{eq:Taylor_Chap}
\mathbb{E} \left[ \frac{n_{10}n_{01}}{n_{11}} \right] \approx \frac{m_{11}m_{10}m_{01}+m_{10}m_{01}}{m_{11}^2} = \frac{m_{10}m_{01}}{m_{11}} \frac{m_{11}+1}{m_{11}}. &&
\end{align}
This implies that $\mathbb{E} \left[ \frac{n_{10}n_{01}}{n_{11}} \right] \frac{m_{11}}{m_{11}+1}$ removes the second-order Taylor approximation bias from the LP-estimator, which suggests that multiplying the LP-estimator with $\frac{n_{11}}{n_{11}+1}$, which gives the Chapman-estimator, is an improvement over the LP-estimator.

\section{Tables with SDs and RMSEs}

\subsection{DSE} \label{App:sim_DSE}

\begin{table}[H]
\setcounter{table}{0}
\renewcommand{\thetable}{C\arabic{table}}
  \caption{The SDs and RMSEs for the simulation study presented in Table \ref{tab:DSE}.} \label{tab:DSE_RMSE}
\begin{tabular}{l|r@{}lr@{}lr@{}lr@{}l|r@{}lr@{}lr@{}lr@{}l}
\toprule
$S$ & \multicolumn{2}{c}{$\text{SD}^{\text{LP}}$} & \multicolumn{2}{c}{$\text{SD}^{\text{Bailey}}$} & \multicolumn{2}{c}{$\text{SD}^{\text{EB/CFK}}$} & \multicolumn{2}{c}{$\text{SD}^{\text{Chap/RL}}$} & \multicolumn{2}{c}{$\text{RMSE}^{\text{LP}}$} & \multicolumn{2}{c}{$\text{RMSE}^{\text{Bailey}}$} & \multicolumn{2}{c}{$\text{RMSE}^{\text{EB/CFK}}$} & \multicolumn{2}{c}{$\text{RMSE}^{\text{Chap/RL}}$} \\
\midrule
1 & 27&.8$^{\dagger}$ & 20&.8 & 25&.8 & 21&.9 & 28&.3$^{\dagger}$ & 21&.2 & 26&.3 & 21&.9 \\
2 & 28&.7 & 22&.2 & 26&.3 & 23&.0 & 29&.3 & 22&.3 & 26&.8 & 23&.0 \\
3 & 70&.2 & 65&.6 & 68&.9 & 66&.5 &70&.7 &65&.9 &69&.3 &66&.5 \\
4 & 85&.7 & 78&.9 & 83&.3 & 79&.7 &86&.5 &79&.0 &83&.8 &79&.7 \\
5 & 460&.9 & 457&.9 & 459&.9 & 458&.4 &461&.2 &458&.1 &460&.1 &458&.4 \\
6 & 411&.3 & 409&.3 & 410&.7 & 409&.6 &411&.7 &409&.3 &410&.9 &409&.6 \\
7 & 109&.6$^{\dagger}$ &45&.8 & 104&.2 &48&.8 & 118&.9$^{\dagger}$ &47&.5 & 107&.9 &49&.4 \\
\bottomrule
\end{tabular}
 \begin{tablenotes}
      \small
      \item       
        A $\dagger$ as superscript indicates that extremely high estimates due to failures were replaced with the highest Chapman estimate in the simulation sample.
 \end{tablenotes}
\end{table}

\subsection{MSE with saturated models} \label{App:sim_saturated}

\begin{table}[H]
\renewcommand{\thetable}{C\arabic{table}}
  \caption{The SDs of the estimates for saturated MSE models, as presented in Table \ref{tab:MSE_saturated}.} \label{tab:MSE_saturated_sd}
\begin{tabular}{l|r|r@{}lr@{}lr@{}lr@{}lr@{}l}
\toprule
$S$ & \multicolumn{1}{r}{$N$} & \multicolumn{2}{c}{$\text{SD}^{\text{ML,SAT}}$} & \multicolumn{2}{c}{$\text{SD}^{\text{EB,SAT}}$} & \multicolumn{2}{c}{$\text{SD}^{\text{CFK,SAT}}$} & \multicolumn{2}{c}{$\text{SD}^{\text{RL,SAT}}$} & \multicolumn{2}{c}{$\text{SD}^{\text{Chap MSE,SAT}}$} \\
\midrule
1 & 100 & 47&.0$^{\dagger}$ & 54&.7 & 40&.2 & 29&.6 & 23&.6 \\
2 & 500 & 103&.9 & 102&.4 & 101&.0 &93&.9 &89&.4 \\
3 & 10,000 & 364&.2 & 364&.1 & 364&.0 & 362&.8 & 362&.1 \\
\midrule
4 & 100 & 58&.5$^{\dagger}$ & 72&.5 & 46&.9 & 33&.3 & 25&.4 \\
5 & 500 & 111&.9 & 109&.6 & 107&.5 &99&.9 &94&.6 \\
6 & 10,000 & 373&.4 & 373&.2 & 373&.0 & 371&.9 & 371&.2 \\
\midrule
7 & 100 & 76&.9$^{\dagger}$ & 86&.4 & 51&.1 & 34&.1 & 24&.7 \\
8 & 500 & 139&.9 & 132&.6 & 126&.8 & 114&.7 & 105&.6 \\
9 & 10,000 & 391&.8 & 391&.6 & 391&.4 & 390&.0 & 389&.1 \\
\midrule
10 & 100 & 66&.5$^{\dagger}$ & 71&.4 & 44&.7 & 30&.5 & 22&.7 \\
11 & 500 & 128&.8$^{\dagger}$ & 134&.3 & 123&.4 & 111&.2 & 103&.0 \\
12 & 10,000 & 394&.5 & 394&.3 & 394&.1 & 392&.7 & 391&.8 \\
\midrule
13 & 20,000 & 636&.7 & 636&.5 & 635&.9 & 635&.4 & 628&.6 \\
14 & 20,000 & 725&.3 & 725&.0 & 724&.4 & 723&.5 & 714&.3 \\
\bottomrule
\end{tabular}
 \begin{tablenotes}
      \small
      \item       
A $\dagger$ as superscript indicates that extremely high estimates due to failures were replaced with the highest Chapman MSE estimate in the simulation sample.
    \end{tablenotes}
\end{table}

\begin{table}[H]
\renewcommand{\thetable}{C\arabic{table}}
  \caption{The RMSEs of the estimates for saturated MSE models, as presented in Table \ref{tab:MSE_saturated}.} \label{tab:MSE_saturated_rmse}
\begin{tabular}{l|r|r@{}lr@{}lr@{}lr@{}lr@{}l}
\toprule
$S$ & \multicolumn{1}{r}{$N$} & \multicolumn{2}{c}{$\text{RMSEs}^{\text{ML,SAT}}$} & \multicolumn{2}{c}{$\text{RMSEs}^{\text{EB,SAT}}$} & \multicolumn{2}{c}{$\text{RMSEs}^{\text{CFK,SAT}}$} & \multicolumn{2}{c}{$\text{RMSEs}^{\text{RL,SAT}}$} & \multicolumn{2}{c}{$\text{RMSEs}^{\text{Chap MSE,SAT}}$} \\
\midrule
1 & 100 & 48&.7$^{\dagger}$ & 56&.1 & 41&.7 & 29&.8 & 23&.6 \\
2 & 500 & 106&.0 & 104&.6 & 103&.2 &94&.1 &89&.4 \\
3 & 10,000 &364&.5 &364&.4 &364&.3 &362&.8 &362&.1 \\
\midrule
4 & 100 & 60&.5$^{\dagger}$ & 73&.9 & 48&.4 & 33&.4 & 25&.4 \\
5 & 500 & 114&.7 & 112&.3 & 110&.1 & 100&.2 &94&.6 \\
6 & 10,000 &373&.7 &373&.5 &373&.4 &371&.9 &371&.2 \\
\midrule
7 & 100 & 79&.8$^{\dagger}$ & 88&.4 & 52&.8 & 34&.3 & 24&.7 \\
8 & 500 & 143&.5 & 136&.1 & 130&.2 & 115&.1 & 105&.6 \\
9 & 10,000 &392&.3 &392&.1 &392&.0 &390&.1 &389&.1 \\
\midrule
10 & 100 & 68&.8$^{\dagger}$ & 73&.1 & 46&.2 & 30&.6 & 22&.8 \\
11 & 500 & 132&.2$^{\dagger}$ & 137&.6 & 126&.9 & 111&.6 & 103&.0 \\
12 & 10,000 &394&.9 &394&.8 &394&.7 &392&.7 &391&.7 \\
\midrule
13 & 20,000 &638&.8 &638&.6 &638&.0 &636&.8 &628&.6 \\
14 & 20,000 &726&.9 &726&.8 &726&.2 &724&.6 &714&.3 \\
\bottomrule
\end{tabular}
 \begin{tablenotes}
      \small
      \item       
A $\dagger$ as superscript indicates that extremely high estimates due to failures were replaced with the highest Chapman MSE estimate in the simulation sample.
    \end{tablenotes}
\end{table}

\subsection{MSE with restricted models} \label{App:sim_restricted}

\begin{table}[H]
\renewcommand{\thetable}{C\arabic{table}}
  \caption{The SDs of the estimates for the correct restricted MSE models, as presented in Table \ref{tab:MSE_restricted}.} \label{tab:MSE_restricted_sd}
\begin{tabular}{l|r|r@{}lr@{}lr@{}lr@{}lr@{}l}
\toprule
$S$ & \multicolumn{1}{r}{$N$} & \multicolumn{2}{c}{$\text{SD}^{\text{ML,LLM$^{\text{sim}}$}}$} & \multicolumn{2}{c}{$\text{SD}^{\text{EB,LLM$^{\text{sim}}$}}$} & \multicolumn{2}{c}{$\text{SD}^{\text{CFK,LLM$^{\text{sim}}$}}$} & \multicolumn{2}{c}{$\text{SD}^{\text{RL,LLM$^{\text{sim}}$}}$} & \multicolumn{2}{c}{$\text{SD}^{\text{Chap MSE,LLM$^{\text{sim}}$}}$} \\
\midrule
1 & 100 & 8&.0 & 7&.9 & 8&.0 & 8&.0 & 7&.8 \\
2 & 500 & 28&.5 & 28&.3 & 28&.5 & 28&.4 & 28&.3 \\
3 & 10,000 & 125&.9 & 125&.8 & 125&.8 & 125&.8 & 125&.8 \\
\midrule
4 & 100 & 11&.7 & 11&.4 & 11&.6 & 10&.9 & 11&.0 \\
5 & 500 & 41&.2 & 40&.7 & 41&.0 & 40&.1 & 40&.3 \\
6 & 10,000 & 164&.4 & 164&.3 & 164&.3 & 164&.2 & 164&.2 \\
\midrule
7 & 100 & 15&.4$^{\dagger}$ & 15&.4 & 14&.3 & 13&.1 & 12&.3 \\
8 & 500 & 48&.3 & 47&.9 & 47&.5 & 46&.6 & 45&.8 \\
9 & 10,000 & 192&.8 & 192&.8 & 192&.7 & 192&.5 & 192&.4 \\
\midrule
13 & 20,000 & 116&.3 & 116&.3 & 116&.3 & 116&.3 & 116&.3 \\
14 & 20,000 & 175&.3 & 175&.3 & 175&.3 & 175&.2 & 175&.2 \\
\bottomrule
\end{tabular}
 \begin{tablenotes}
      \small
      \item       
        A $\dagger$ as superscript indicates that extremely high estimates due to failures were replaced with the highest Chapman estimate in the simulation sample.
 \end{tablenotes}
\end{table}

\begin{table}[H]
\renewcommand{\thetable}{C\arabic{table}}
  \caption{The RMSEs of the estimates for the correct restricted MSE models, as presented in Table \ref{tab:MSE_restricted}.} \label{tab:MSE_restricted_rmse}
\begin{tabular}{l|r|r@{}lr@{}lr@{}lr@{}lr@{}l}
\toprule
$S$ & \multicolumn{1}{r}{$N$} & \multicolumn{2}{c}{$\text{RMSE}^{\text{ML,LLM$^{\text{sim}}$}}$} & \multicolumn{2}{c}{$\text{RMSE}^{\text{EB,LLM$^{\text{sim}}$}}$} & \multicolumn{2}{c}{$\text{RMSE}^{\text{CFK,LLM$^{\text{sim}}$}}$} & \multicolumn{2}{c}{$\text{RMSE}^{\text{RL,LLM$^{\text{sim}}$}}$} & \multicolumn{2}{c}{$\text{RMSE}^{\text{Chap MSE,LLM$^{\text{sim}}$}}$} \\
\midrule
1 & 100 &8&.0 &8&.0 &8&.1 &8&.0 &7&.8 \\
2 & 500 &28&.6 &28&.3 &28&.6 &28&.4 &28&.3 \\
3 & 10,000 &125&.9 &125&.8 &125&.9 &125&.8 &125&.8 \\
\midrule
4 & 100 & 11&.8 & 11&.4 & 11&.8 & 10&.9 & 11&.0 \\
5 & 500 &41&.4 &40&.8 &41&.2 &40&.1 &40&.3 \\
6 & 10,000 &164&.4 &164&.3 &164&.4 &164&.2 &164&.2 \\
\midrule
7 & 100 & 15&.7$^{\dagger}$ & 15&.7 & 14&.6 & 13&.2 & 12&.3 \\
8 & 500 &48&.6 &48&.2 &47&.8 &46&.6 &45&.8 \\
9 & 10,000 &192&.9 &192&.8 &192&.8 &192&.6 &192&.4 \\
\midrule
13 & 20,000 &116&.3 &116&.3 &116&.3 &116&.3 &116&.3 \\
14 & 20,000 &175&.3 &175&.3 &175&.3 &175&.2 &175&.2 \\
\bottomrule
\end{tabular}
 \begin{tablenotes}
      \small
      \item       
        A $\dagger$ as superscript indicates that extremely high estimates due to failures were replaced with the highest Chapman estimate in the simulation sample.
 \end{tablenotes}
\end{table}

\end{document}